\documentclass[apj]{emulateapj}

\usepackage{graphicx}
\usepackage{amssymb}
\usepackage{amsmath}
\usepackage{natbib}

\def\kmsec{\mbox{km~s$^{\rm -1}$}}
\def\logg{\mbox{log~{\it g}}}

\def\teff{\mbox{$T_{\rm eff}$}}

\def\loggf{\mbox{$\log gf$}}

\shorttitle{Expanded Sample of Red Giants in the LMC Bar}
\shortauthors{SONG ET AL.}

\begin{document}
\title{An Expanded Chemo-dynamical Sample of Red Giants in the Bar of the Large Magellanic Cloud}

\author{Ying-Yi\ Song}
\affiliation{Department of Astronomy, University of Michigan, 1085 S. University Ave., Ann Arbor, MI 48109, USA; yysong@umich.edu}
\author{Mario\ Mateo}
\affiliation{Department of Astronomy, University of Michigan, 1085 S. University Ave., Ann Arbor, MI 48109, USA}
\author{Matthew G.\ Walker}
\affiliation{McWilliams Center for Cosmology, Department of Physics, Carnegie Mellon University, 5000 Forbes Ave., Pittsburgh, PA 15213, USA}
\author{and Ian U.\ Roederer}
\affiliation{Department of Astronomy, University of Michigan, 1085 S. University Ave., Ann Arbor, MI 48109, USA}
\affiliation{Joint Institute for Nuclear Astrophysics and Center for the Evolution of the Elements (JINA-CEE), USA}

\begin{abstract}
We report new spectroscopic observations obtained with the Michigan/{\em Magellan} Fiber System of 308 red giants (RGs) located in two fields near the photometric center of the bar of the Large Magellanic Cloud.  This sample consists of 131 stars observed in previous studies (in one field) and 177 newly-observed stars (in the second field) selected specifically to more reliably establish the metallicity and age distributions of the bar.  For each star, we measure its heliocentric line-of-sight velocity, surface gravity and metallicity from its high-resolution spectrum (effective temperatures come from photometric colors).  The spectroscopic Hertzsprung-Russell diagrams---modulo small offsets in surface gravities---reveal good agreement with model isochrones.  The mean metallicity of the 177-RG sample is $\rm [Fe/H]=-0.76\pm0.02$ with a metallicity dispersion $\sigma=0.28\pm0.03$.  The corresponding metallicity distribution---corrected for selection effects---is well fitted by two Gaussian components: one metal-rich with a mean $-0.66\pm0.02$ and a standard deviation $0.17\pm0.01$, and the other metal-poor with $-1.20\pm0.24$ and $0.41\pm0.06$. The metal-rich and metal-poor populations contain approximately 85\% and 15\% of stars, respectively.  We also confirm the velocity dispersion in the bar center decreases significantly from $31.2\pm4.3$ to $18.7\pm1.9$~km~s$^{-1}$ with increasing metallicity over the range $-2.09$ to $-0.38$.  Individual stellar masses are estimated using the spectroscopic surface gravities and the known luminosities.  We find that lower mass hence older RGs have larger metallicity dispersion and lower mean metallicity than the higher-mass, younger RGs.  The estimated masses, however, extend to implausibly low values ($\rm \sim 0.1~M_{\odot}$) making it impossible to obtain an absolute age-metallicity or age distribution of the bar.
\end{abstract}


\section{Introduction}
\label{sec:intro}
The Large Magellanic Cloud (LMC) is the nearest gas-rich satellite of
the Milky Way that has ongoing star formation. The visual structure
of the LMC is dominated by a prominent central bar, and hence we
define the galaxies of this kind as Barred Magellanic Sprials
\citep{dV+F72} or SB(s)m \citep{1991RC3.9}. Magellanic systems are
common in the local universe, but they are rarely found as close to a
massive parent system as the LMC to the Milky Way \citep{Wilcots04,
  Liu11, Tollerud11, Robotham12}.  Because of its relative proximity
and prominence, the central bar of the LMC represents a particularly
useful test case to explore how such structures develop and evolve in
galactic systems.

It is well known that the LMC has a stellar bar that has no
counterpart in the distribution of neutral or ionized gas
\citep[e.g.,][]{Kim98, Bica99, SS03}.  More interestingly, the stellar
bar is found to have multiple centers.  The photometric center of the
LMC bar, i.e. the densest point on a star count map, lies about 0.4
kpc away from the center of its stellar disk, and even more than 1 kpc
away from the dynamical center of the neutral gas disk \citep{vdM01}.
The location of the stellar dynamical center is still under debate.
According to the line-of-sight (LOS) kinematics, the dynamical center
of the carbon stars is consistent with the photometric bar center
\citep{vdM02, Olsen11}.  In contrast, the average proper motion (PM)
data observed by \textit{Hubble Space Telescope} (\textit{HST}) imply
the stellar dynamical center coincides with the H~\textsc{i} dynamical
center \citep{Kallivayalil06, Piatek08, Kallivayalil13, vdM14}.  When
combining the LOS velocities and the PM data, \citet{vdM14} even found
that the stellar dynamical center is intermediate between the
photometric bar center and the H~\textsc{i} dynamical center.

\citet{KK04} distinguish `true' bars, which are formed in a
quasi-independent manner early in the history of a disk galaxy, from
`secular' bars that develop over time from instabilities in disk
galaxies.  The off-center feature of the LMC bar strongly supports the
secular evolution scenario resulting from tidal interactions.  Many
numerical simulations have been employed to reproduce the morphology
and internal dynamics of the LMC including its off-center bar, but the
detailed evolution remains uncertain.  For example, \citet{Bekki09}
showed that the off-center bar can be formed if the LMC with an
already-existing bar can collide with a low-mass Galactic subhalo.
Alternatively, more recent works prefer a dwarf-dwarf galaxy
interaction \citep[e.g.,][]{Besla12, Yozin14, Pardy16}, presumably, in
the case of the LMC, involving the Small Magellanic Cloud (SMC).

The star clusters in the LMC are rarely found to have ages between
approximately 3 and 12 Gyr ago \citep[e.g.,][]{Olszewski96,
  Geisler97}, though this age interval appears to be filled by field
stars \citep[e.g.,][]{Holtzman99, SH02}.  This suggests either that
the star formation associated with star clusters was suppressed during
this `age gap', or that clusters older than about 4 Gyr are
preferentially destroyed (though not all; the LMC does contain a set
of about ten `ancient' star clusters comparable in age to globular
clusters in the Galaxy).  Whatever the cause for the age gap in the
cluster population, the fact remains that if we aim to probe the
history of the LMC for ages greater than about 4 Gyr, we must rely on
field stars.  Although many photometric studies have been carried out
to do this \citep[e.g.,][]{SH02}, interpretation is complicated by the
increased insensitivity of the main-sequence turnoff with age, and the
fact that all evolved stars older than about 1 Gyr essentially funnel
into a single red-giant branch for a given metallicity
\citep{Bressan12}.  In the LMC bar, crowding introduces an additional
complication for ground-based studies, while HST observations tend to
have limited field coverage.  Spectroscopic studies of the LMC bar can
be used as as a tool to determine the age-metallicity relation of the
bar and to explore the star formation history of this enigmatic
component of the LMC.

Many spectroscopic studies in the LMC bar and/or inner disk has been
carried out during the past two decades.  \citet{Gratton04} analyzed
low resolution spectroscopy of 98 RR Lyrae stars in the LMC bar, and
reported an average metallicity of $-1.48$ dex for this old
population.  \citet[][hereafter `C05']{Cole05} carried
out the first spectroscopic study of intermediate-age and old field
stars in the LMC bar.  The spectra of 373 red giants (RGs) were
obtained at the near-infrared Ca~\textsc{ii} triplet (CaT) and used to
derive radial velocities and metallicities.  They found a metallicity
distribution function (MDF) peaked at $\rm [Fe/H]=-0.40$ dex with a
tail of metal-poor stars down to $\rm [Fe/H]\leq-2.1$ dex and a
systemic change in velocity dispersion with mean chemical abundance.
\citet{Carrera11} reanalyzed C05's spectra and also
obtained an average metallicity of $-0.40$ dex with a new calibration.
On the other hand, studies on the inner disk reported lower mean
metallicities than that of C05.  \citet{Carrera08}
derived $\rm [Fe/H]\approx-0.5$ dex for a field about $3^\circ$ north
of the LMC bar at first, and then \citet{Carrera11} recalibrated a new
mean of $-0.58$ dex for the same field.  \citet{Lapenna12} derived
$\rm [Fe/H]=-0.48$ dex from 89 stars located about $2^\circ$ NW from
the center of the LMC.  More recently, \citet[][hereafter
  `VdS13']{VdS13} performed a detailed chemical analysis of 106 RGs in
the sample of C05, using spectra obtained with the
FLAMES/GIRAFFE multifibre spectrograph.  Their measurements confirmed
that C05 had overestimated the metallicities of
metal-rich stars (by $0.25\pm0.03$~dex from our calculation).  Unfortunately, a reliable MDF of
the LMC bar cannot be derived from their results due to inherent
metallicity-dependent biases in their sample.

Since the nature of LMC bar remains poorly constrained in terms of the
galaxy's interaction and chemical evolution history, we aim to expand
previous spectroscopic studies by observing a new sample of stars
chosen using well-defined selection criteria.  The ultimate aim is to
produce a spectroscopic survey of evolved stars that can be used to
map the chemo-dynamical properties as a function of position and age
over the entire LMC bar region.  In addition to 131 RGs observed by
C05, our sample contains 177 more RGs selected from
the OGLE-II photometry database \citep{Udalski97, Udalski00, Szymanski05}
for fully providing the bar MDF.  The spectra were obtained
with the Michigan/\textit{Magellan} Fiber System \citep[M2FS,
][]{Mateo12} over two separate fields that both are near the
photometric center of the LMC bar.  This paper represents a first look at
the results of this survey and provides a description
of the techniques we employ.

The structure of the paper is as follows.  Section \ref{sec:obs&data}
introduces the sample selection, the observations and the data
reduction processes.  Section \ref{sec:fit} explains the measurements of
velocities and stellar parameters that we derive from our spectra.
Section \ref{sec:results} reports the main results of this work and
describes some of their implications.  In Section \ref{sec:discussion}, we
summarize and further discuss the key results of this study.  We close
with a cautionary tale regarding the inherent and still significant
complications in using field stars to independently probe the age
distribution/star-formation history in an intermediate-age/old
population such as the LMC bar.

\section{Observations and Data Reduction}
\label{sec:obs&data}
\subsection{Fields and Target Selection}
\label{sec:selection}
\begin{figure*}[htbp]
   \centering
   \includegraphics[width=0.75\textwidth]{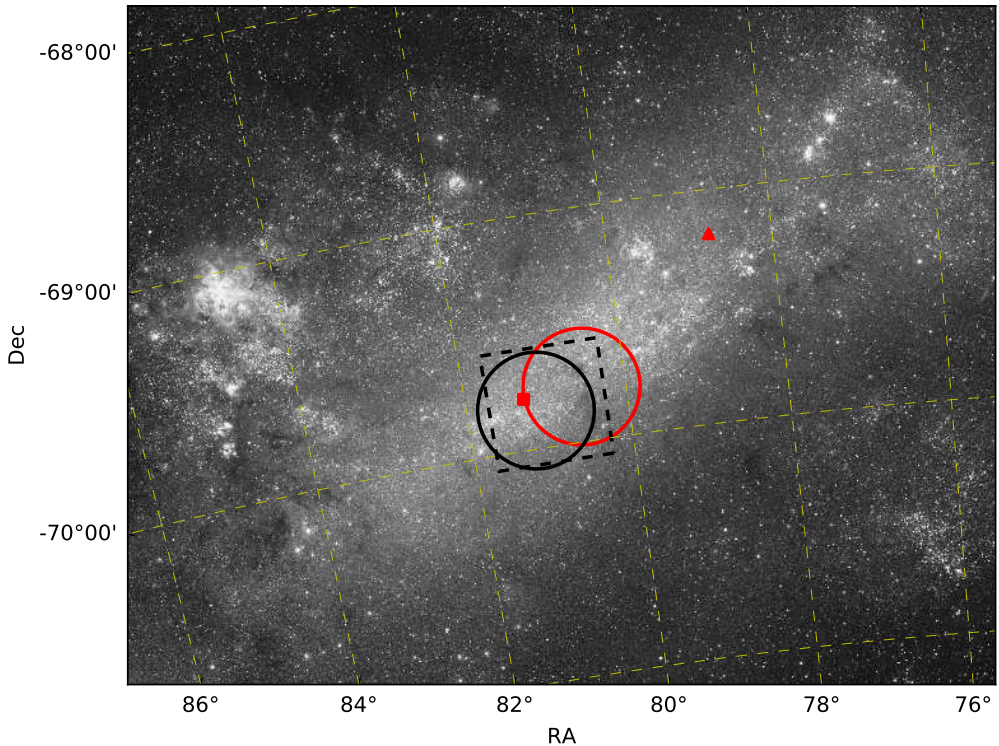}
   \caption{
   Schematic map of the LMC bar, showing the locations of our observed fields (solid circles) relative to the bar and other prominent features in the LMC. 
   The dashed rectangle shows the region containing seven FORS2 pointings from C05 (see their Figure 2). 
   The spectroscopic targets reported in this work are located in the LMC1 (red circle) and LMCC (black circle) fields.
   The red filled square shows the densest point in the LMC bar as defined by \citet{vdM01}, and the red filled triangle represents the stellar dynamical center inferred from a model fit to the third-epoch \emph{HST} proper motions by \citet{vdM14}.  
   }
   \label{fig:map}
\end{figure*}

In standard operation, M2FS fields must be centered on a relatively
bright `Shack-Hartmann' (SH) star, which provides low-frequency
wavefront data to the active optics system of the
\textit{Magellan}/Clay Telescope.  Two additional spatially-coherent
fiber bundles are used to image a pair of guide stars during
exposures.  Finally, a set of bright acquisition stars imaged through
science fibers and visible on the guide camera are used for each field
to held centroid the fibers in the field in both translation and
rotation.  These requirements ancillary stars impose mild restrictions
on any M2FS field locations, but especially in crowded regions like
the LMC bar.  The data obtained for this study were collected in two
bar fields labelled as `LMCC' and `LMC1', respectively.  The trailing
letter of `LMCC' stands for `Cole' because this field was chosen to
include as many stars as practical from the sample of
C05.  The centers of these fields are listed in
Table \ref{tab:obs}, and their locations and coverage areas are shown in
Figure \ref{fig:map} along with the photometric bar center and the
dynamical center constrained by stellar proper motions \citep{vdM14}.

We used two different methods to select the stellar candidates in the
LMCC and LMC1 fields. In LMCC, the candidate RGs were directly
selected from the sample of C05 by first ranking their stars by metallicity and then selecting every other
one to produce a suitably-sized subsample for M2FS followup.  In LMC1,
the candidate RGs were randomly selected from the OGLE-II $BVI$ maps
of the LMC \citep{Udalski00}, according to the following photometric
criteria:
\begin{equation}
\label{eq:criteria}
16.00< I < 17.00,\ 22.00 < I + 5\times(V-I) < 24.25.
\end{equation}
These limits were chosen to sample fully the color range--and hence
the evidently wide metallicity range--of the population.  As shown in
Figure \ref{fig:CMD_selection}, we further divided the selection region
into 32 rectangles and selected 6 candidates from each.  This approach
ensures that we sampled stars over a full color/metallicity/age range
as populated within the red giant branch and allows us to account for
selection when we generate the underlying metallicity distribution in
the bar (see Section \ref{sec:MDF}).

As already intimated, the operational characteristics of M2FS affect
the selection process.  M2FS employs aluminum fiber plugplates to
position up to 256 optical fibers at the Nasmyth-East focal surface of
the \textit{Magellan}/Clay Telescope.  Each fiber has an entrance
aperture of 1.2 arcsec and is fitted in a ferrule 13 arcsec in
diameter; the latter defines the minimum separation between deployed
fibers.  Otherwise, any fiber can be positioned anywhere within a
field of 29 arcmin in diameter except at locations for various
ancillary stars used for guiding, field alignment/acquisition, and
active optical control (the central SH star noted above).  In the end,
we were able to assign 147 science fibers in the LMCC field, and 184 science fibers in LMC1.
Table \ref{tab:sample} lists the positions and photometric information
of the observed stars. We also plot the locations of both LMCC and
LMC1 samples on the color-magnitude diagram (CMD) in Figure \ref{fig:CMD}.

\subsection{Spectroscopic Observations}
\label{sec:obs}
The summary of our LMC bar observations is listed in Table \ref{tab:obs}
including the total exposure time for each field.  M2FS employs twin
spectrographs (which are referred to as `blue' and `red' channels,
respectively) that have identical optical properties and wavelength
coverage but can be operated independently in a variety of spectral
configurations.  In our LMCC and LMC1 observations, both spectrographs
were configured to sample a wavelength range of 5130--5189 \AA
at an effective resolution $\mathcal{R} \sim 20,000$.  Each
spectrograph images the raw spectra onto a four-channel $4096\times
4112$ E2V CCD with a pixel size of $\rm 15~\mu m$.  During the readout
process, we binned the data by $2\times2$, which still over-samples
the data in both spectral and spatial directions. The CCDs were
readout in `slow' mode for a typical gain of 0.75 e$^-$/ADU and a
readout noise of 2.7 e$^-$.

Immediately before or after science exposures, we acquired calibration
frames from a Th-Ar arc lamp (for wavelength calibration) and a quartz
lamp (for spectral tracing).  In addition, several twilight frames
(which also included arc and quartz exposures) were obtained at either
the beginning or the end of the same observing night.  These twilight
data are used to check the wavelength/velocity calibration and correct
fiber-to-fiber throughput variations (Section \ref{sec:reduction}), and
also to estimate systematic offsets in the best-fit physical
parameters (Section \ref{sec:offsets}).  We also obtained a master `fiber
map' with all the fibers plugged in while the telescope focal surface
is illuminated by ambient daylight in the mostly-closed dome.  The map
produces a high signal-to-noise template for tracing spectra, and is
useful as a parallel check on the relative fiber throughputs estimated
from the twilights.  Groups of bias (zero) and dark frames (also by
$2\times2$ binning) were also obtained during each observing run and
combined to produce master bias and dark frames that are used in the
data reduction.

Background subtraction is very important in this work because of the
crowded LMC bar region.  The central surface brightness of the LMC bar
(see the red filled square in Figure \ref{fig:map}) is $20.65\ \rm
mag/arcsec^2$ in $V$-band \citep{Bothun88}.  In contrast, the $V$-band
telluric sky background would have ranged from about 22.0 (Dec 2014)
to 21.6 (Nov 2015) $\rm mag/arcsec^2$ at the location of the LMC
bar---about 3.5--2.4 times fainter than the LMC contribution---at the
time of our observations.

We have therefore adopted two distinct approaches to obtain the
background spectra in the LMCC and LMC1 fields, respectively.  In
LMCC, 95 sky-background fibers were assigned in addition to the 147 science fibers, and their positions were selected on the image
obtained by STScI DSS.  For selecting good background positions, we
first randomly picked a position within the LMCC field and kept it if
meeting two conditions: (a) the chosen position is well separated ($>
13$ arcsec) from all other assigned fiber positions, and (b) the mean
count of all pixels within a $5\time5$ square centered on that
position is within 10\% of the modal background value of the image.  In
LMC1, no sky fibers were assigned in advance due to the initial use of
these data for more limited M2FS commissioning purposes.  
To estimate the background contribution, we pointed the telescope 15 arcsec
away from the original position---known as `off-target' exposures---along three principal directions (North, South and East) during the Nov 2015 observation run. 
As a result, the background subtraction approaches differ slightly for the LMCC and LMC1 samples; we describe these in the following section.

\subsection{Data Reduction}
\label{sec:reduction}
Condensed descriptions of standard M2FS data reduction processes can
be found in \citet{Walker15a, Walker15b}.  All data processing was
carried out using IRAF scripts and pipelines designed for M2FS data,
and the typical final products are background-subtracted stellar
spectra.

Briefly, all data were first processed through overscan subtraction,
bias correction and dark correction.  We removed cosmic rays from
almost all exposures (excluding arcs or other short exposures) using
the Laplacian-filtering algorithm from \citet{vD01}.  We subtracted
diffuse scattered light from the two-dimensional images by fitting a
polynomial surface to the regions of the images not illuminated by the
fibers.  The spectral traces, defined by combining
twilights/quartzes/fiber maps, were shifted to match the locations of
the science spectra (named because they are extracted from the science
exposures) that produced final science traces.  The final science
traces were used to extract the calibration arcs to ensure no offset
or interpolation shift existing between the arc exposures and the
science exposures.  The extracted arcs were then fit to a
moderate-order polynomial to determine the transformation from
extracted-pixel to wavelength for every science fiber/target.  The
typical root-mean-square of these fits for the data used in this study
was 0.3 km~s$^{-1}$.  The wavelength-calibrated data were then
normalized using relative fiber throughputs derived from the twilights
or fiber maps.  For all the above steps, we processed the sky-background
data---both from the sky-background fibers in LMCC and from the
off-target exposures in LMC1---the same as the science data.
Throughout the entire reduction process, we also calculated `variance
spectra' in order to track the signal-to-noise ratio (SNR) for every
pixel of every science spectrum.  These data are used to properly
weight individual pixels when fitting spectra to model atmospheres in
order to derive stellar parameters (see Section \ref{sec:Bayesian}).

The last but important step is the background subtraction.  In LMCC,
we found no significant flux or spectral variations among
sky-background fibers.  We therefore averaged all these background
spectra to make a master background spectrum and subtracted it
directly from all science spectra.  In LMC1, the background spectra
were obtained from the off-target exposures one year later.  After
removing some anomalously high-flux spectra (about 10\% of the total),
we averaged all other spectra to made a second-year master background
spectrum. Then this spectrum was scaled by a factor 1.8 before being
subtracted from the deeper first-year science spectra.  This factor
accounts for differences in exposure times, atmospheric extinction,
telluric background and seeing between the two sets of observations.
The systematic reliability of this approach is confirmed in Section \ref{sec:MDF} where we demonstrate good agreement in the metallicity scales
obtained independently from the M2FS data in the LMCC and LMC1 fields.

Figure \ref{fig:spec} shows examples of background-subtracted spectra in
the LMCC and LMC1 fields.  The spectra span a range from the highest
to lowest SNRs for our targets, and also a range in
metallicities (measured through the method introduced in the following section).

\section{Analysis}
\label{sec:fit}
\subsection{Modeling of M2FS Spectra}
\label{sec:Bayesian}

To measure the LOS velocity and the stellar parameters, we employ a
Bayesian method to fit the background-subtracted M2FS spectrum
(see \citet{Walker15a, Walker15b}).  For each star, this approach
generates a model spectrum $M(\lambda)$ by combining a
continuum-normalized template spectrum $T(\lambda)$ and an assumed
continuum spectrum $P_{l}(\lambda)$,
\begin{equation}
M(\lambda)=P_l(\lambda)T(\lambda),
\end{equation}
where $P_l(\lambda)$ is an $l$-order polynomial.  This model spectrum
$M(\lambda)$ is used to compare with the observed spectrum using a
maximum-likelihood technique.

The template spectrum, $T(\lambda)$, is generated using a library of
synthetic spectra, which is used in the Sloan Extension for Galactic
Exploration and Understanding (SEGUE) stellar Parameter Pipeline
\citep[SSPP, ][]{Lee08a, Lee08b}.  This library contains a set of
rest-frame, continuum-normalized, stellar spectra, which are computed
over a grid containing three atmospheric parameters, i.e. the effective temperature ($T_{\rm eff}$), the surface gravity ($\log{g}$,
where $g$ is in $\rm cm\,s^{-2}$) and the metallicity ($\rm
[Fe/H]$). These parameters vary in the following ranges (with the grid
steps):
\begin{eqnarray}
4000 \leq& T_{\rm eff} &\leq 10000\ {\rm K}, {\rm with}\ \Delta T_{\rm eff}=250\ {\rm K}, \nonumber \\
0 \leq& \log{g} &\leq 5~{\rm dex}, \ {\rm with}\ \Delta \log{g}=0.25\ {\rm dex}, \\
-5 \leq &{\rm [Fe/H]}& \leq 1~{\rm dex}, \ {\rm with}\ \Delta {\rm [Fe/H]}=0.25\ {\rm dex}, \nonumber
\end{eqnarray}
For the $\alpha$ elements, their total abundance is typically fixed based on the
iron abundance via the ratio $\rm [\alpha/Fe]$. This library
assumes a hard-wired relation between $\rm [\alpha/Fe]$ and $\rm
[Fe/H]$:
\begin{equation}
  {\rm [\alpha/Fe]} =
  \begin{cases}
    0.4 & \text{for } {\rm [Fe/H]} < -1, \\
    -0.4\times {\rm [Fe/H]} & \text{for } {\rm [Fe/H]} \in \left[-1,\,0\right], \\
    0 & \text{for } {\rm [Fe/H]} \geq 0.
  \end{cases}
\end{equation}

Another consideration is the wavelength shift between the template
spectrum and the observed spectrum.  This shift has two potential
origins: the velocity shift due to the LOS velocity and the
uncertainty of the wavelength solution in the observed spectrum. The
first offset is accounted for by a parameter that shifts the
wavelength as $\lambda' = \lambda v_{\rm los}/c$, where $v_{\rm los}$
is the LOS velocity and $c$ is the speed of light.  The wavelength
solution may have residual systematic deviations, which could be
modeled as a polynomial $Q_m(\lambda)$ of order $m$. 
Combining these two effects, the final shifted wavelength is
\begin{equation}
\lambda' = \lambda \left[ 1+\frac{Q_m(\lambda)+v_{\rm los}}{c} \right],
\end{equation}
and so the final template spectrum is now represented as
$T(\lambda')$.

To carry out the maximum-likelihood technique, we adopt the likelihood
function used by \citet{Walker15a},
\begin{eqnarray}
\label{eq:likelihood}
&&\mathcal{L} (S(\lambda)|\vec{\theta},\,s_1,\,s_2) = \nonumber \\
&&  \prod_{i=1}^{N_\lambda} \frac{1}{\sqrt{2\pi(s_1 {\rm Var} [S(\lambda_i)]+s_2^2)}} 
\exp \left[ -\frac{1}{2} \frac{\left( S(\lambda_i) - M(\lambda_i) \right)^2}{s_1{\rm Var}[S(\lambda_i)]+s_2^2}\right], \nonumber \\
\end{eqnarray}
where $S(\lambda_i)$ and $M(\lambda_i)$ are the observed spectrum and
the model spectrum, respectively, and $\vec{\theta}$ is a vector of
all free parameters in $M(\lambda_i)$ (summarized in the following paragraph).  There are
also two nuisance parameters $s_1$ and $s_2$ that, respectively,
rescale and add an offset to the observational variances to account
for systematically misestimated noises.

In practice, $P_l(\lambda)$ is treated as a fifth order polynomial
(i.e., $l=5$) incorporating six parameters.  The wavelength-shifted
$T(\lambda')$ inherits three parameters from the library spectra and
three coefficients related to the modification of $Q_m(\lambda)$
(i.e., $m=2$).  An additional parameter is varied to convolve the
template spectra with the instrumental line-spread function (LSF), and
thus $\vec{\theta}$ ends up with 13 free parameters.  Considering the
two nuisance parameters $s_1$ and $s_2$ in Eq. \ref{eq:likelihood}, the full fitting
method contains 15 free parameters in the end, among which four of
them are the physical parameters we aim to measure: $v_{\rm los}$,
$T_{\rm eff}$, $\log{g}$ and $\rm [Fe/H]$.  In each fitting, we
truncate the spectrum in the region $5130 \leq \lambda /$\AA$\leq 5180$, while for the template spectra, the rest-frame wavelength
region $5120 \leq \lambda /$\AA$\leq 5190$ are adopted to
account for the LOS velocities up to about $\pm 550\rm\ km\ s^{-1}$.

\subsection{Priors of the Physical Parameters}
\label{sec:priors}
We use the MultiNest package \citep{MultiNest08, MultiNest09,
  MultiNest13} to scan the parameter space.  MultiNest implements a
nested-sampling Monte Carlo algorithm, and returns random samplings
from the posterior probability distribution functions (PDFs) for all
input parameters. We record the first four moments of each physical parameter's posterior
PDF: mean, variance, skewness and kurtosis.

MultiNest requests a set of prior distributions for all parameters to
initiate the calculation. We adopted
uniform priors over a specified range of values as listed in Table~2
of \citet{Walker15a}.  For the physical parameters $v_{\rm los}$,
$\log{g}$ and ${\rm [Fe/H]}$, the priors were set within the ranges:
\begin{eqnarray}
&-500 \leq v_{\rm los}/({\rm km\ s^{-1}}) \leq 500, \nonumber \\
&0 \leq \log{[g/({\rm cm\ s^{-2}})]} \leq 5, \\
&-5 \leq {\rm [Fe/H]}/{\rm dex} \leq 1. \nonumber
\end{eqnarray}

Different from \citet{Walker15a}, $T_{\rm eff}$ was fixed during the fitting process in this study, because it proved difficult to constrain the temperatures
adequately given the narrow wavelength range of our spectra and the
typically modest median SNR per pixel (ranging from 5 to around 25).
In addition, we found that the best-fit values for $\log{g}$ and $\rm
[Fe/H]$ were strongly correlated with $T_{\rm eff}$ when using uniform
temperature priors.  To break this degeneracy, we calculated effective
temperatures for our targets from the OGLE-II $V-I$ color index using
color-temperature relation for giants \citep[Equation~1 of][]{RM05b}.
We adopted a single-extinction model of $E(V-I)=0.15\pm0.07$ mag,
which is equivalent to $E(B-V)\approx0.11$ mag, $A_V\approx0.34$ mag
and $A_I\approx0.20$ mag in the UBVRI photometric system.  Since the
color-temperature relation reported by \citet{RM05b} is a weak
function of metallicity, we simply assumed $\rm [Fe/H]=-0.8$ dex, the
mean value for our targets when we used the Bayesian method with
$T_{\rm eff}$ as a free parameter.

\subsection{Twilight Offsets and Errors}
\label{sec:offsets}
We also applied the Bayesian method described above to the twilight
spectra obtained on the same night and in the same spectrograph
configuration.  Using the solar effective
temperature $T_{\rm eff,\,\odot}=5778\ {\rm K}$ as the fixed temperature
prior, we found that the parameters $v_{\rm los}$, $\log{g}$ and $\rm
[Fe/H]$ fitted from those spectra deviate from the known solar values,
$v_{\rm los,\,\odot}=0\ {\rm km\,s^{-1}}$, $\log_{10}{[g_{\odot}/(\rm
    cm\,s^{-2})]}=4.44$ dex and $\rm [Fe/H]_{\odot}=0$ dex.  As
discussed in \citet{Walker15a, Walker15b}, we attribute this to
systematic mismatches between the model spectrum and the observed
spectra.  Without independent information regarding how such
mismatches may vary with spectral type, we defined the offsets between
the fitting and the true twilight parameters to be zero-point
shifts. Table \ref{tab:twi_offsets} lists the mean offsets and the
standard deviations of all three fitting physical parameters in each
field and each channel.

According to Table \ref{tab:twi_offsets}, we apply the twilight offsets in $\rm [Fe/H]$ to all
results obtained from the Bayesian analysis of our science spectra. 
For $v_{\rm los}$ and $\log{g}$, we chose not to apply offsets to the data
as the required offsets are either not statistically significant, or
they have no significant implications for our final results.
The total error budget for our
derived stellar parameters include contributions from the variances of
the twilight offsets, the variances of the posterior PDFs of $v_{\rm
  los}$, $\log{g}$ and $\rm [Fe/H]$, and---for $T_{\rm eff}$---the
uncertainty of the color excess $\sigma[E(V-I)]=0.07$~mag. 

A full table of the final results is available in machine-readable
format online.  The initial few lines of this table is provided in
Table \ref{tab:param}.  We list only results for spectra with median SNR per pixel
greater than 5 ($\rm SNR>5$), which results in 133 (out of 147
observed) targets in LMCC and 179 (out of 184) in LMC1.  Excluding
double/blended stars (see Section \ref{sec:HRD}), the final LMCC and LMC1
sample sizes are 131 and 177, respectively. 
Heliocentric corrections has been applied to all velocity
results and were calculated using appropriate
PyAstronomy\footnote{https://github.com/sczesla/PyAstronomy} routines.

\subsection{Metallicity from the Equivalent Widths}
\label{sec:EW}

In order to examine the accuracy of the Bayesian analysis (BA), we
compare our metallicities with those derived using the traditional
spectroscopic analysis.  For this purpose, we selected 11 LMCC stars
(see Table \ref{tab:comp}) that not only have high-SNR spectra ($\rm
SNR>10$) but also show significant differences in the measured
metallicities between our work and VdS13
($\rm|[FeI/H]_{VdS13}-[Fe/H]_{Bayesian}|>0.2$~dex).

We restrict the analysis to Fe~\textsc{i} lines with accurate
\loggf\ values (grade ``D$+$'' or better, $\sigma <$~0.22~dex,
according to the NIST Atomic Spectra Database; \citealt{kramida15}),
which leaves us with three lines that can be measured in our spectra
(5150.84, 5166.28 and 5171.60~\AA).  Our approach is to measure
equivalent widths (EWs) of these lines through a semi-automated
routine that fits Voigt line profiles to continuum-normalized spectra.
We inspect these fits and then modify them if the automated routine
clearly fails to identify the continuum, which often occurs when the
routine incorrectly includes neighboring absorption lines in the
calculation of the continuum.  The uncertainties in the EW values are
$\approx$~10\%.

Before applying this EW method to our M2FS spectra, we first measure
EWs of these lines in high-quality spectra of two metal-poor RG
standard stars, Arcturus \citep{hinkle00} and \mbox{HD~175305}
\citep{roederer14}.  The metallicities derived for the two standard
stars are in good agreement with previous work. For Arcturus, we
derive [Fe/H]~$= -$0.41~$\pm$~0.11 dex, which agrees with the value
[Fe/H]~$= -$0.52~$\pm$~0.04 dex derived by \citet{ramirez11}. For
\mbox{HD~175305}, we derive [Fe/H]~$= -$1.61~$\pm$~0.34 dex, which
agrees with the value [Fe/H]~$= -$1.56~$\pm$~0.15 dex derived by
\citet{roederer14}. This indicates that for high resolution and high
SNR stellar spectra, our EW method is reliable, and the choice of
lines has little effect on the derived metallicities.

As an additional test, we have also degraded the resolution and SNR of these spectra to
match our typical M2FS spectra. After remeasuring the EWs, the derived [Fe/H] values for these two standard stars are found to be
approximately 0.2~dex lower than when the high-quality spectra are used.  The statistical
uncertainties of the derived abundances, however, are quite large (about 0.4~dex), so
we do not consider this to represent strong evidence of a systematic offset in [Fe/H]
induced by the modest SNR of our spectra.   Finally, we also applied the BA to degraded
spectra of Arcturus and HD~175305.  In each case, we obtain [Fe/H] results to within 0.2~dex
of the `true' values.

For these 11 selected stars, we use the newly measured EW values with the formerly measured
\teff\ and \logg\  (listed in Table \ref{tab:comp}) to derive the metallicity for each star. 
We adopt a constant microturbulence velocity parameter, $v_t = 2.0 \pm 0.4$~\kmsec, for each
star because we lack any constraints on this parameter.
We derive the metallicities using a recent version of the spectrum analysis code
MOOG \citep{sneden73,sobeck11}. The calculations are repeated 250
times for each star, resampling the stellar parameters and EWs each
time from normal distributions. 

Table \ref{tab:comp} lists the metallicities derived by our BA and EW methods,
as well as those measured by VdS13 and C05.
We calculated the weighted mean differences in the metallicities among those measurements as 
\begin{eqnarray}
&&\langle \rm [Fe/H]_{\rm BA}- [Fe/H]_{\rm EW}\rangle = 0.07 \pm 0.12~\rm dex, \nonumber \\
&&\langle \rm [Fe/H]_{\rm V13}- [Fe/H]_{\rm EW}\rangle= 0.22 \pm 0.14~\rm dex, \\
&&\langle \rm [Fe/H]_{\rm C05}- [Fe/H]_{\rm EW}\rangle= 0.25 \pm 0.15~\rm dex. \nonumber 
\end{eqnarray}
At face value, this analysis favors the BA results, but there are important caveats.
For example, the limited wavelength coverage provides few reliable Fe~\textsc{i} lines for the
EW analysis, so the statistical uncertainties are large (see Table \ref{tab:comp}).  The relatively
low SNR of the spectra may also mask the presence of other weak lines in the spectra leading to
a systematically low misplacement of the continuum level. This interpretation is supported by
the results of our test to rederive [Fe/H] from the degraded standard star spectra
described above. Nonetheless, lacking a set of standard stars observed in the same conditions for calibration, we
find no need for any additional corrections to the BA metallicities beyond the twilight
offset described in Section \ref{sec:offsets}.   We shall therefore adopt the BA metallicities as
listed in Table \ref{tab:param} throughout the rest of this paper. 

\subsection{Comparison with Previous Studies}
As stated in Section \ref{sec:offsets}, we have 133 RGs left in the LMCC sample that were selected from C05's sample.  Among them, 39 were also reobserved by VdS13. 
These stars provide an opportunity to compare the physical parameter measurements on a star-by-star basis.
In Figure \ref{fig:comp}, we compare the LOS velocities and the stellar parameters measured from our M2FS spectra
to those reported by C05 (top panels) and VdS13 (middle and bottom panels). 

For the subsample of 133 stars, we measured a mean heliocentric LOS velocity
of $262.9 \pm 2.1$~km~s$^{-1}$ and a corresponding velocity dispersion of $24.4\pm1.6$~km~s$^{-1}$,
compared to $258.7 \pm 2.4$~km~s$^{-1}$ and $27.0\pm1.7$~km~s$^{-1}$ measured by C05 (Figure \ref{fig:comp}a). 
The median errors of the individual velocities are $0.4$~km~s$^{-1}$ (this work) and $7.5$~km~s$^{-1}$ (C05).
There is a systematic offset of $4.2\pm0.8$~km~s$^{-1}$ in the velocity measurements between these two studies (in the sense M2FS$-$C05). 
Figure \ref{fig:comp}b illustrates that we typically measure lower metallicities than C05 did for stars in common. 
Excluding the two anomalous sources in the upper-left corner of Figure \ref{fig:comp}b (see Section \ref{sec:HRD}),
the remaining 131 stars reveal weighted mean metallicities of $-0.76 \pm 0.03$~dex (this work) and
$-0.50 \pm 0.03$~dex (C05).  The corresponding metallicity dispersions are
$0.30\pm0.03$~dex and $0.37\pm0.04$~dex, respectively.
Although the dispersions are statistically equivalent, the systematic
metallicity offset of $0.27\pm0.04$~dex between our work and C05
appears to be significant.

For the subsample of 39 stars, we measured a mean heliocentric LOS velocity of
$263.9\pm3.6$~km~s$^{-1}$, which is slightly greater than $262.4\pm3.6$~km~s$^{-1}$ by
VdS13 and $258.9\pm3.9$~km~s$^{-1}$ by C05 (Figure \ref{fig:comp}c). 
The velocity dispersion is $22.0\pm2.4$~km~s$^{-1}$ from our measurements,
compared to $22.1\pm2.5$~km~s$^{-1}$ (VdS13) and $24.3\pm2.7$~km~s$^{-1}$ (C05). 
The median errors in the single-star velocities are $0.3$~km~s$^{-1}$ (our work), $0.2$~km~s$^{-1}$ (VdS13), and $7.5$~km~s$^{-1}$ (C05). 
The systematic offset between our velocities and VdS13 is $1.5\pm0.2$ km~s$^{-1}$ (in the sense M2FS$-$VdS13).

For the same 39 stars, we measured a mean metallicity value of [Fe/H] $= -0.80\pm0.05$~dex, compared to
$-0.71\pm0.06$~dex from VdS13, and $-0.56\pm0.06$~dex from C05. 
Given the median errors on metallicity in three studies are $0.06$ dex (this work), $0.12$ dex (VdS13)
and $0.13$ dex (C05), we confirm that C05 seems to have
overestimated the
metallicity for metal-rich stars in the LMC bar by around $0.25$~dex.  There is less evidence that such an
offset exists among more metal-poor stars ($\rm [Fe/H] < -1.0$~dex) (\citealt{Pompeia08}; VdS13), though the number
of such stars we can directly compare from the various studies is not large.  In contrast, 
our metallicity measurements are in statistically good agreement with those of VdS13.

Finally, we also compared \teff\ and \logg\ between our work and VdS13 in Figures 5e and 5f,
respectively. Although VdS13 and we used different color-\teff\ relations to
calculate \teff\ from photometry, the results of \teff\ are in reasonably good agreement.
The comparison of \logg\ shows more scatter with no clear pattern.
Note that VdS13 calculated their \logg\ from photometry (see Section 5.1),
while our \logg\ are measured by comparing the scientific spectra to a library of templates.
Our results imply poor agreement between these methods, which is an important conclusion for our aim to measure
the age distribution function of LMC bar stars directly (Section 5.1).

\section{Results}
\label{sec:results}

\subsection{Spectroscopic Hertzsprung-Russell Diagram}
\label{sec:HRD}
It is well known from stellar evolutionary models \citep[e.g.,][]{{Girardi00, Bressan12}} and
observations of star clusters \citep[e.g.,][]{Mermilliod81, Meynet93} that the classical red giant branch (RGB) in
the Hertzsprung-Russell (HR) diagram 
typically consists of stars older than approximately
1 Gyr over a wide range of metallicities.
The resulting age-metallicity degeneracy along the RGB  stems from this evolutionary
funneling effect \citep[as it was called decades ago; e.g., ][]{Eggen71}, as well as the competing
effects of age and metallicity on the photometric colors of such stars.
From an astrophysical standpoint, this degeneracy makes it difficult to
disentangle the age or metallicity distribution of an intermediate-age or old RG population
from photometric observations.  
A key aim of this study is to use spectroscopy to try to break this degeneracy as far as
possible by providing independent information on RGB  metallicities and ages (via \logg\ measurements).

To illustrate the potential of this approach, we plot four spectroscopic HR diagrams showing the relationship between the stars' surface gravities and their effective temperatures in
Figure \ref{fig:HRD}.
On each diagram, the stars are color-coded by the same metallicity bins shown in the legend of Figure \ref{fig:HRD}a.
The first three diagrams (Figures 6a, 6b and 6c) are made to illustrate the complications in interpreting the LMCC sample.
We reduce the LMCC sample size from 133 (in Figure \ref{fig:HRD}a) to 62 (in
Figure \ref{fig:HRD}c) for the purpose of investigating the `peculiar' metal-rich stars shown in the upper-left corner of Figure \ref{fig:HRD}a.
No such star has been found in the LMC1 sample and Figure \ref{fig:HRD}d shows a clear segregation by metallicity.

At a first glance, the metal-dependent pattern in Figure \ref{fig:HRD}d
is very similar to that shown on the CMD of C05 (see
their Figure 8).  However, there is a significant difference between
those two diagrams.  In Figure \ref{fig:HRD}d, the color-coded
populations can be well separated by metallicity, with the extremely
metal-poor stars located on the upper-left corner and the extremely
metal-rich ones on the lower-right.  On the other hand, the extremely
metal-poor stars on C05's CMD were overlapped by many
relatively metal-rich counterparts, since the bluest stars in their
sample are mostly metal-rich and a board color range is covered by the
stars with metallicity around the peak of their MDF ($\rm
-0.6<[Fe/H]<-0.3~dex$).

\begin{figure*}[htbp]
   \centering
   \includegraphics[width=0.75\textwidth]{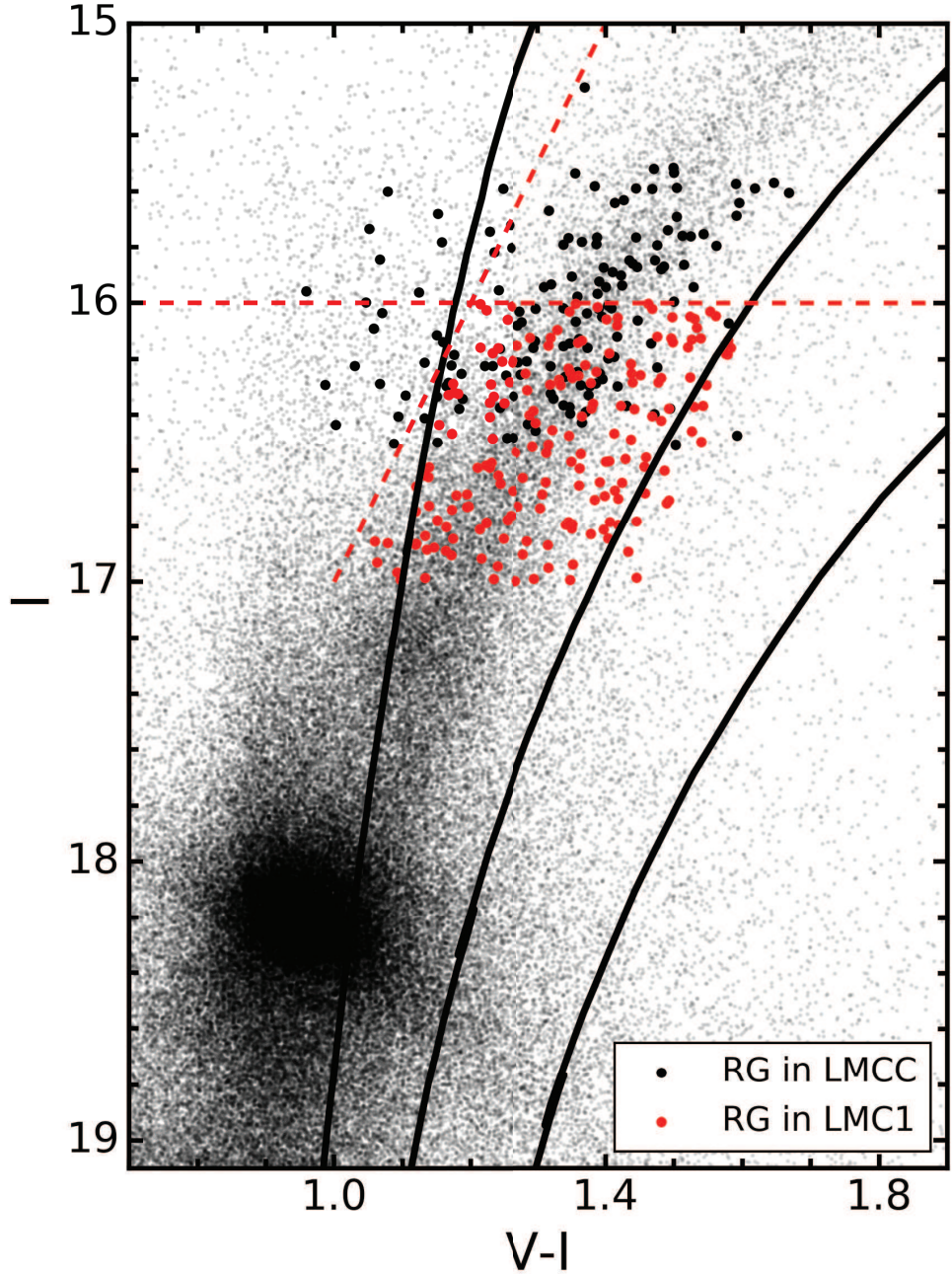}
   \caption{The color-magnitude diagram of the LMC bar with our selected RG targets highlighted for each field shown in Figure \ref{fig:map}. The smaller black dots represent the $V$ and $I$ band photometry from OGLE-II catalogue \citep{Udalski00} of all stellar candidates in the LMC1 field defined Figure \ref{fig:map}. The larger dots represent the targets we selected for spectroscopic observation:  black for the LMCC sample and red for the LMC1 sample (see Figure \ref{fig:map}).
Note that the LMCC targets were selected from the transformed CTIO photometry according to C05, while the LMC1 ones were selected from OGLE-II photometry directly. 
The two red dashed lines show the upper and left boundaries for our LMC1 selection criteria on the CMD. Black curves are the isochrones from \citet{Bressan12} for an age of 6 Gyr and $\rm [Fe/H]=-2.2$ (left), $-0.6$ (middle) and 0.0 (right) dex, respectively. The isochrones are plotted assuming $E(V-I)=0.15$ mag and $A_I=0.20$ mag.}
   \label{fig:CMD}
\end{figure*}

As described in Section \ref{sec:selection}, the stars in our LMCC sample
were selected exclusively from C05's sample.  Their
selection criteria on the CMD (also reflected by the large black dots
in Figure \ref{fig:CMD}) were chosen to include stars over a wide range
of metallicity expected in the LMC bar.  However, the blue edge of
their selection region---chosen to include metal-poor RGs---is likely
contaminated by metal-rich stars younger than about 1 Gyr, which arise
either from the bar itself or from a superimposed disk population in
the central region of the LMC (see the isochrones in Figure \ref{fig:CMD}
as an example).  Such stars show up clearly in Figure \ref{fig:HRD}a,
populating in the region of high $T_{\rm eff}$ and low $\log{g}$
(i.e. upper-left corner of the plot).

When imposing the same blue color cut as we did for LMC1 (i.e. $I +
5\times(V-I)>22.00$ according to Eq. \ref{eq:criteria} and also see
the tilt red dashed line in Figure \ref{fig:CMD}), we exclusively remove
these more massive giants, resulting in a sample of 109 stars shown in
Figure \ref{fig:HRD}b.  Though a few old metal-poor stars are removed in
this selection, the resulting HRD cannot be distinguished from that of
the LMC1 sample (Figure \ref{fig:HRD}d). C05's sample
also extends to a higher luminosity than our LMC1 sample (see the
horizontal red dashed line in Figure \ref{fig:CMD}).  So we select
another subsample from the LMCC sample using the same limits for the
LMC1 sample (Eq. \ref{eq:criteria}). 
The remaining 62 stars in Figure \ref{fig:HRD}c produce an HRD in good
agreement with the HRD of the full LMC1 sample shown in
Figure \ref{fig:HRD}d.

We note that a few apparently very metal-poor stars are located in the lower-right corner of both Figures 6b and 6d, where we would expect the most metal-rich giants.  
In the nomenclature of Table \ref{tab:param}, their IDs are LMCC-b086, LMCC-r001, LMC1-b017 and LMC1-b080.  
On close inspection, we found that the spectra of LMC1-b017 and LMC1-b080 can be well fitted with a two-star rather than a single-star model.
In Figure \ref{fig:spec_double}, their observed spectra are plotted on the left with their best-fit single-star model spectra.  
On the right, the best-fit spectra employing a two-star model is shown for comparison.  
For LMC1-b017, the reduced chi-square value decreases from 1.7 (single-star) to 1.0 (two-star); and for LMC1-b080, the reduced chi-square value decreases from 5.2 (single-star) to 2.0 (two-star).
In both cases, the two-star fits are significantly better.

These two stars appear to be either spectroscopic binaries or physically
unrelated stars that happen to be blended photometrically.  We favor
the second interpretation because double-RG binaries should be
quite rare on stellar evolutionary timing grounds.  For the two cases
shown in Figure \ref{fig:spec_double}, the relative velocity shifts of the individual stars
are in $40.0\pm0.9$~km~s$^{-1}$ for LMC1-b017 and $81.5\pm0.4$~km~s$^{-1}$ for
LMC1-b080.
Such velocities would be very difficult to explain in binary systems
consisting of RGs with masses around 1 $\rm M_\odot$.  
Spectral blending---whether due to a physical or photometric binary---can also
explain the inferred low metallicities of the stars.  Since the stars
have different systemic velocities, their lines will appear weaker
relative to the approximately doubled continuum, resulting in a low
inferred metallicity when fit as a single star.  
The other two anomalous stars in Figure \ref{fig:HRD}b, i.e. LMCC-b086 and LMCC-r001, do not exhibit obviously composite spectra and it is unclear why they exhibit relatively high surface gravities for their low metallicities. 
Alternatively, these stars may be metal-poor subgiants located foreground to the LMC in the Galactic halo.
Whatever the origin, we exclude these four stars from our RG sample and subsequent analysis.  
This resulting LMCC and LMC1 samples consist of 131 and 177 stars, respectively. 

Finally, we compare the results of our reduced LMC1 sample to the
predictions of stellar models from \citet{Bressan12} in Figure \ref{fig:HRD_iso}.  
We applied a systematic offset of $-0.3$~dex to the \logg\ scale of the stellar models to better match the data in a systematic sense.  
Given this shift, the overall trends and ranges of \logg\ and $T_{\rm eff}$ expected in the models appears to fit the data reasonably well.  
Matching these isochrones represents a practical way to constrain the ages of individual RGs if their masses are well known.  We discuss the limitations of
such mass calculations in Section \ref{sec:mass}.

\subsection{Metallicity Distribution Function}
\label{sec:MDF}
Figure \ref{fig:MDF} shows a set of MDFs derived from our reduced LMCC and LMC1 samples.  In all cases, we have fit the
observed MDFs (the histograms) with two Gaussian
distributions (the curves; also known as the two Gaussian mixture model), each of which represents a population
parameterized by a Gaussian mixture weight $w_i$, a mean value $\mu_i$
and a standard deviation $\sigma_i$.  The Gaussian mixture weight is
the fraction of a single Gaussian distribution relative to the sum of
the two Gaussians, and all weights should sum to unity.  The best-fit
Gaussian parameters can be found in the legend of each panel in
Figure \ref{fig:MDF}.

C05 first fitted the MDF of the LMC bar by two
populations. Their original MDF contains 373 RGs and can be best
fitted by two Gaussian distributions as
$(w_1,\,\mu_1,\,\sigma_1)=(0.89,\,-0.37,\,0.15)$ and
$(w_2,\,\mu_2,\,\sigma_2)=(0.11,\,-1.08,\,0.46)$.  One goal of our
LMCC selection was to acquire a subsample from C05
without changing their original MDF.  It can be confirmed---by the
dashed histogram in Figure \ref{fig:MDF}a with its best-fitting
curve---that there is no significant difference between the MDF of
the original C05 sample and the MDF of 131 stars.  
With this subsample of 131 stars, we also confirm that the fitting results of
two Gaussian distributions are the best among one to ten Gaussian
mixture models, and so is the case for all the following MDFs.
Figure \ref{fig:MDF}a also provides a direct comparison between the MDFs
of the same 131 RGs from C05's (dashed) and our (solid) metallicity measurements.  Our MDF peaks at a lower
metallicity for both the metal-rich and metal-poor populations; the
differences relative to C05 are $\rm 0.27~dex\ and\ 0.14~dex$ (in the sense C05$-$M2FS), respectively.

In Figures 9b and 9c, we compare the MDFs between the LMCC and LMC1
samples.  The LMC1 MDF remains the same in both figures, with the
best-fit parameters as
$(w_1,\,\mu_1,\,\sigma_1)=(0.85\pm0.06,\,-0.69\pm0.02,\,0.16\pm0.01)$
and
$(w_2,\,\mu_2,\,\sigma_2)=(0.15\pm0.06,\,-1.23\pm0.20,\,0.40\pm0.05)$.
The LMCC MDFs, on the other hand, are changed due to the different
sample sizes: all 131 stars in the reduced LMCC sample are included in
Figure \ref{fig:MDF}b; while in Figure \ref{fig:MDF}c, only the 62 stars are considered for being located in the same CMD region determined by the LMC1 sample (see
Section \ref{sec:HRD} and Figure \ref{fig:HRD}c).  Both LMCC MDFs agree well
with the LMC1 MDF.

\begin{figure*}[htbp]
   \centering
   \includegraphics[width=0.75\textwidth]{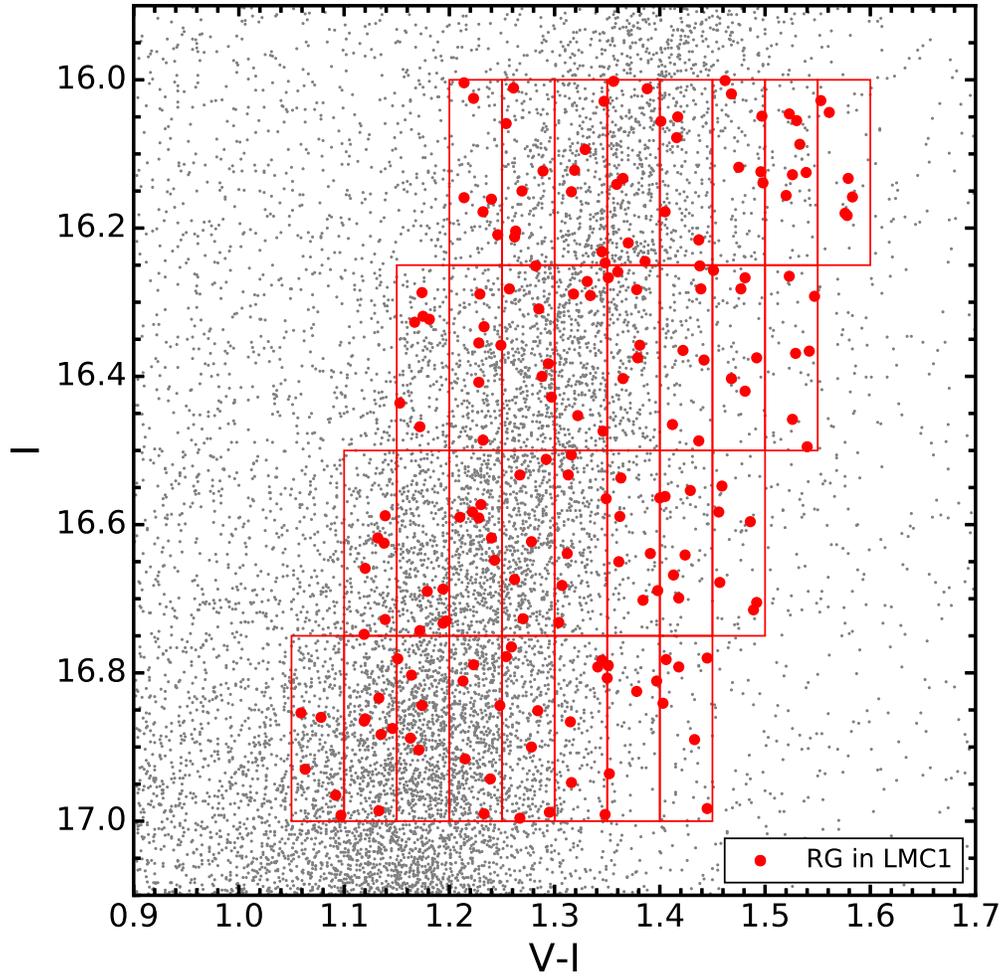}
   \caption{The RG selection criteria in the LMC1 field. The small black dots represent the OGLE-II photometry of all stars in this field. Six candidates were selected within each red rectangle. }
   \label{fig:CMD_selection}
\end{figure*}

\begin{figure*}[htbp]
   \centering
   \includegraphics[width=0.49\textwidth]{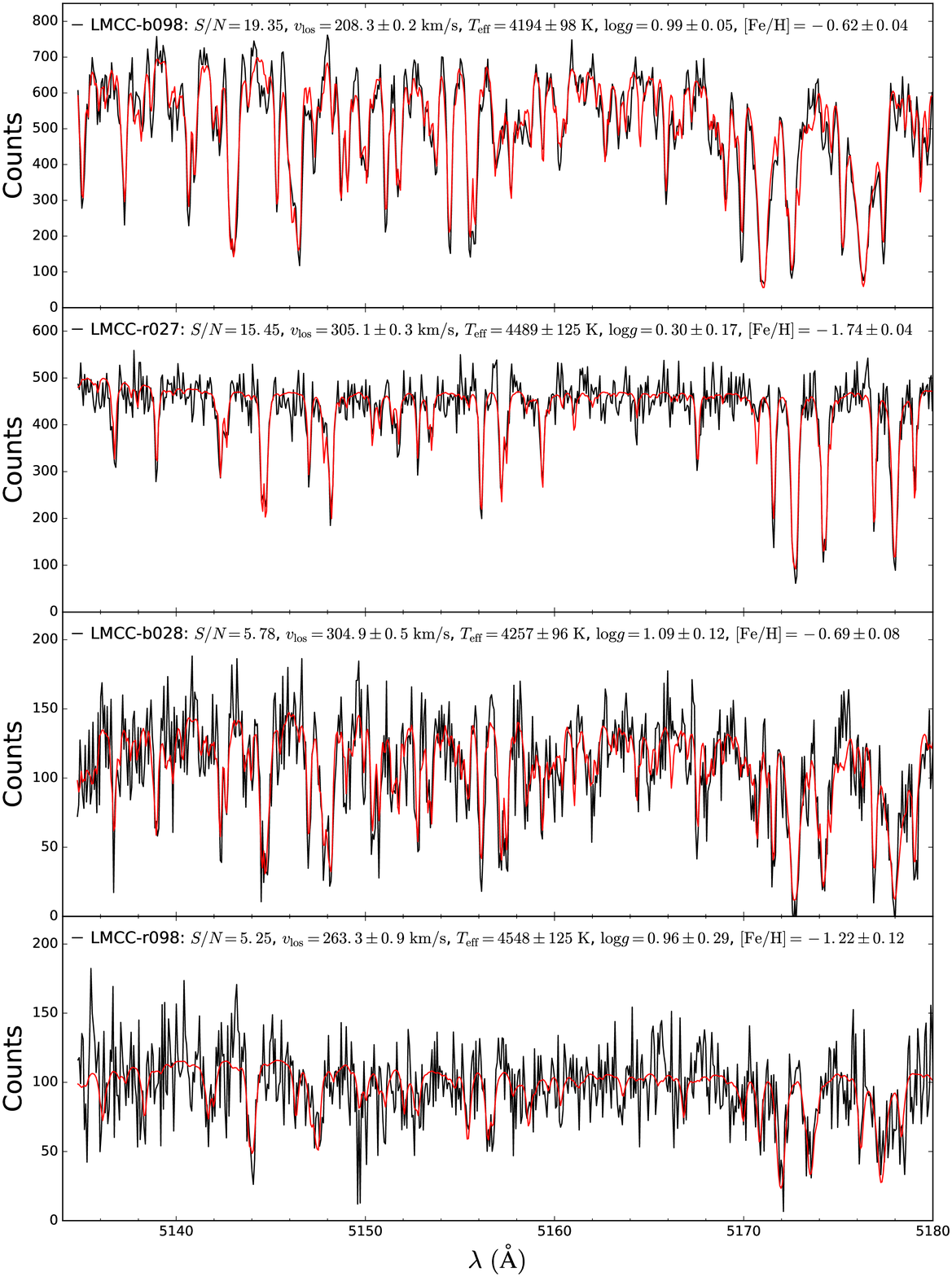}
   \includegraphics[width=0.49\textwidth]{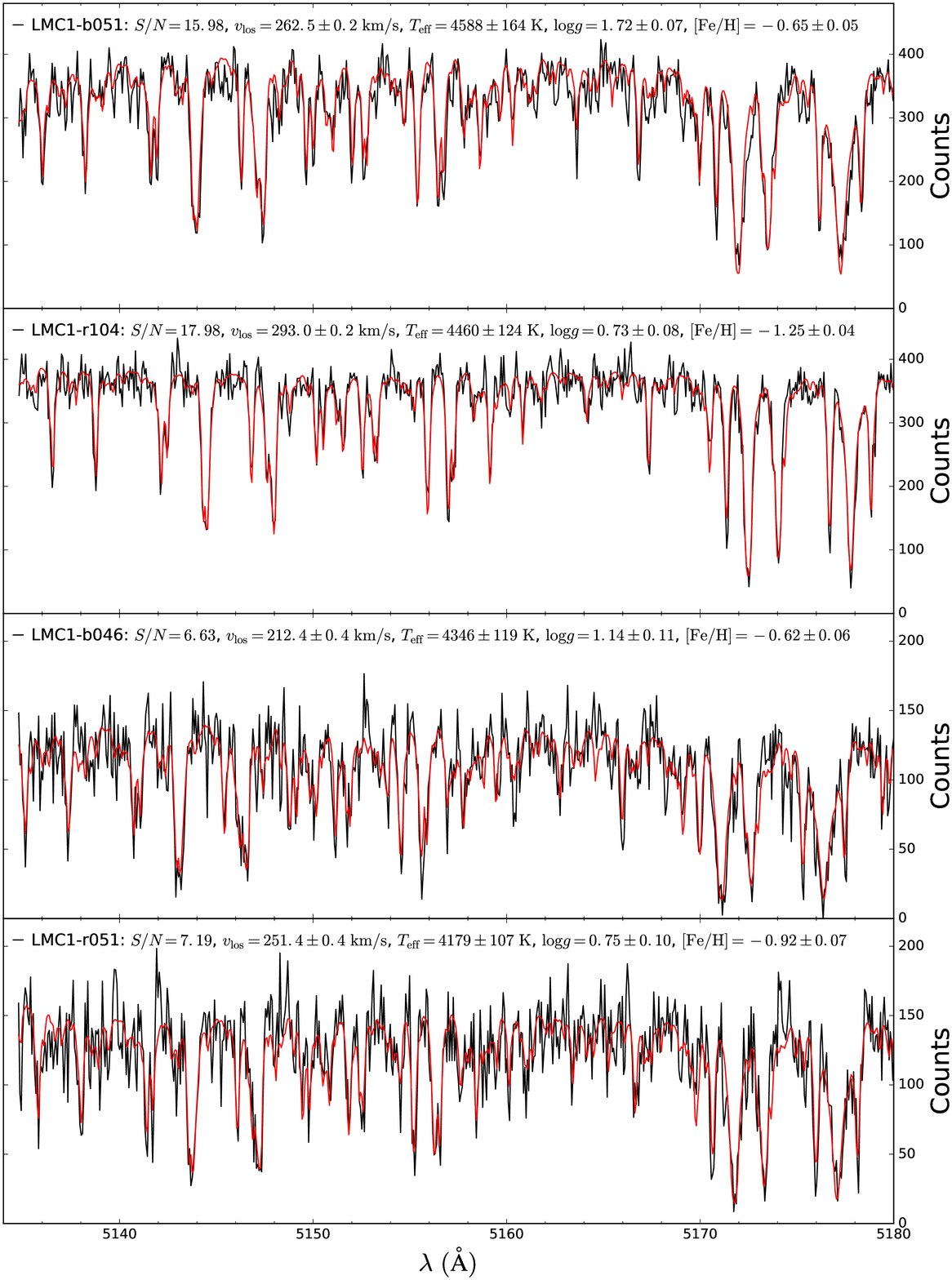}
   \caption{A sample of selected background-subtracted M2FS spectra (black) in the LMCC (left panels) and LMC1 (right panels) fields, with best-fitting models (red). Text lists target ID, median SNR per pixel, and our measurements of $v_{\rm los}$, $T_{\rm eff}$, $\log g$, and ${\rm [Fe/H]}$. In both fields, we selected the sample spectra of high-SNR metal-rich (top panel), high-SNR metal-poor (second panel), low-SNR metal-rich (third panel) and low-SNR metal-poor (bottom panel), respectively. }
   \label{fig:spec}
\end{figure*}

\begin{figure*}[htbp]
   \centering
      \includegraphics[width=0.4\textwidth]{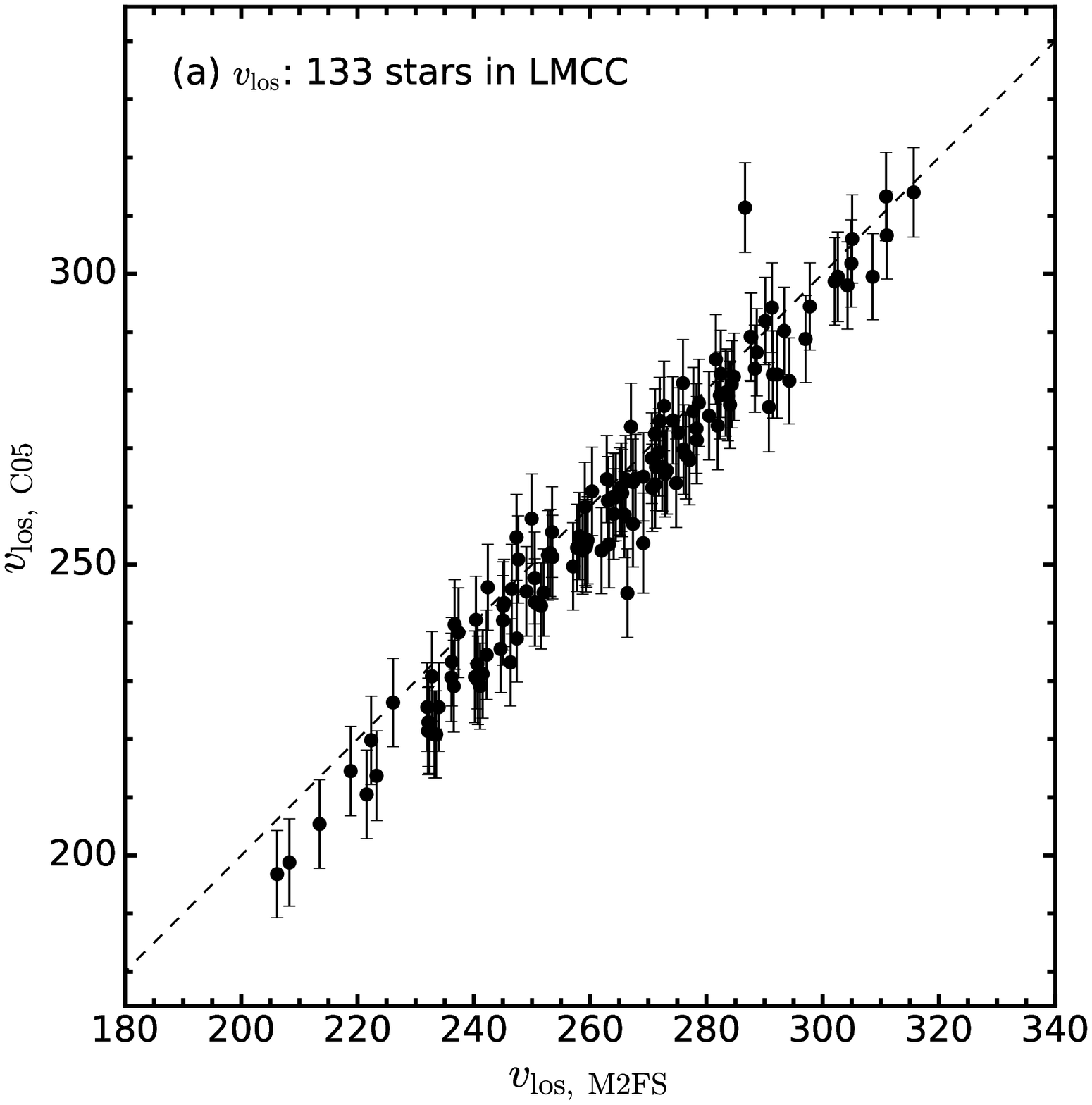}
      \includegraphics[width=0.4\textwidth]{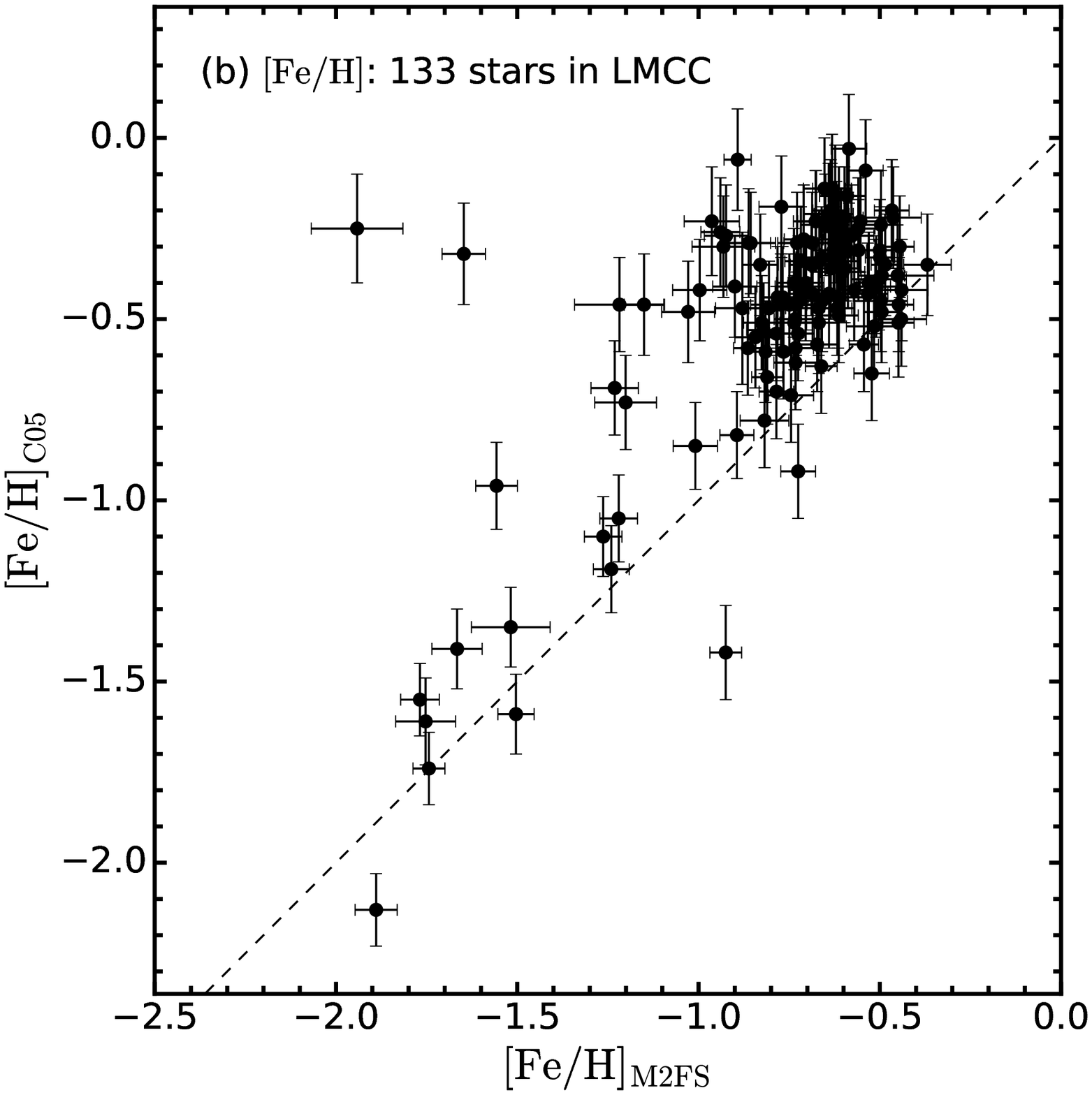}
      \includegraphics[width=0.4\textwidth]{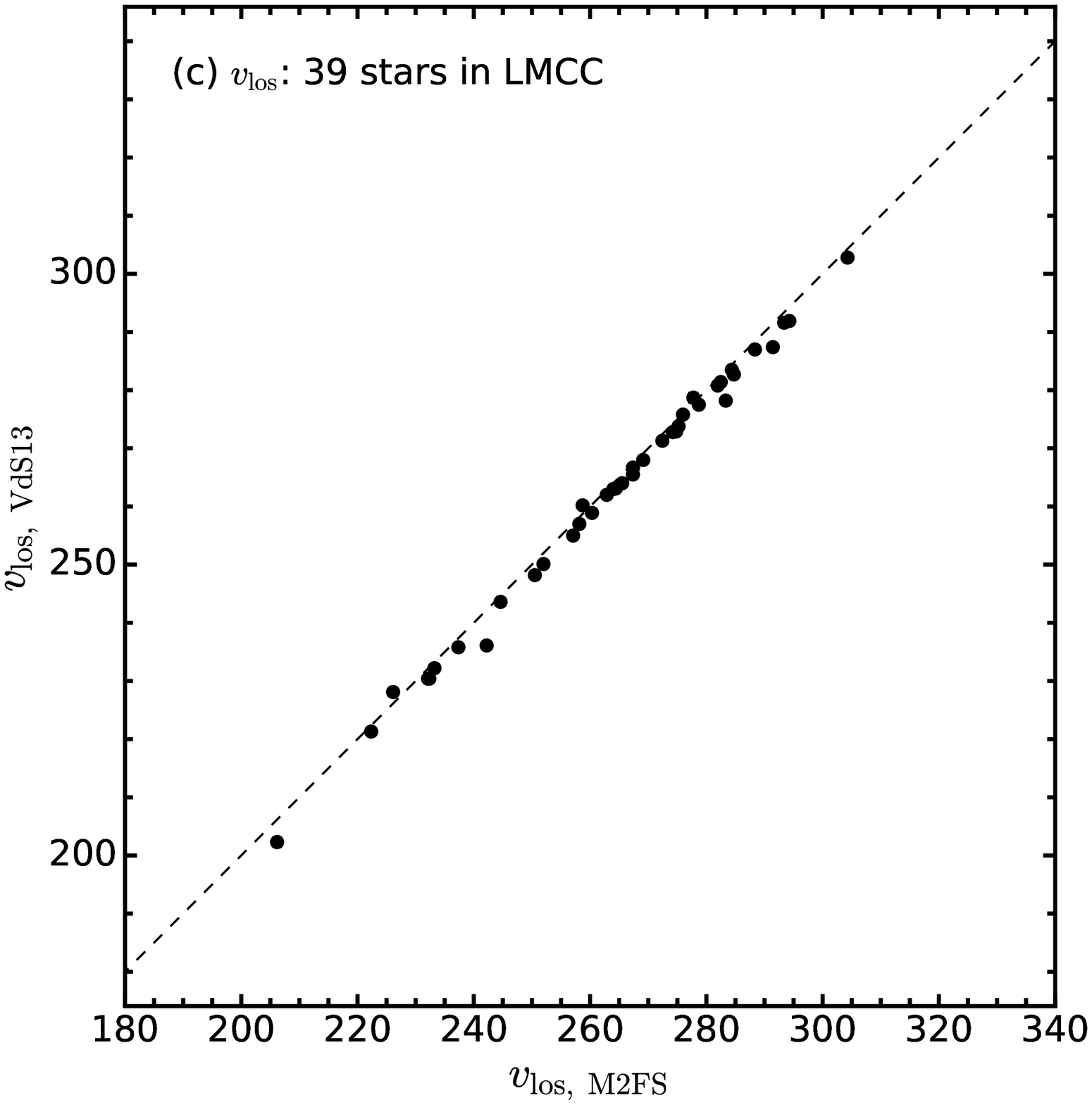}
      \includegraphics[width=0.4\textwidth]{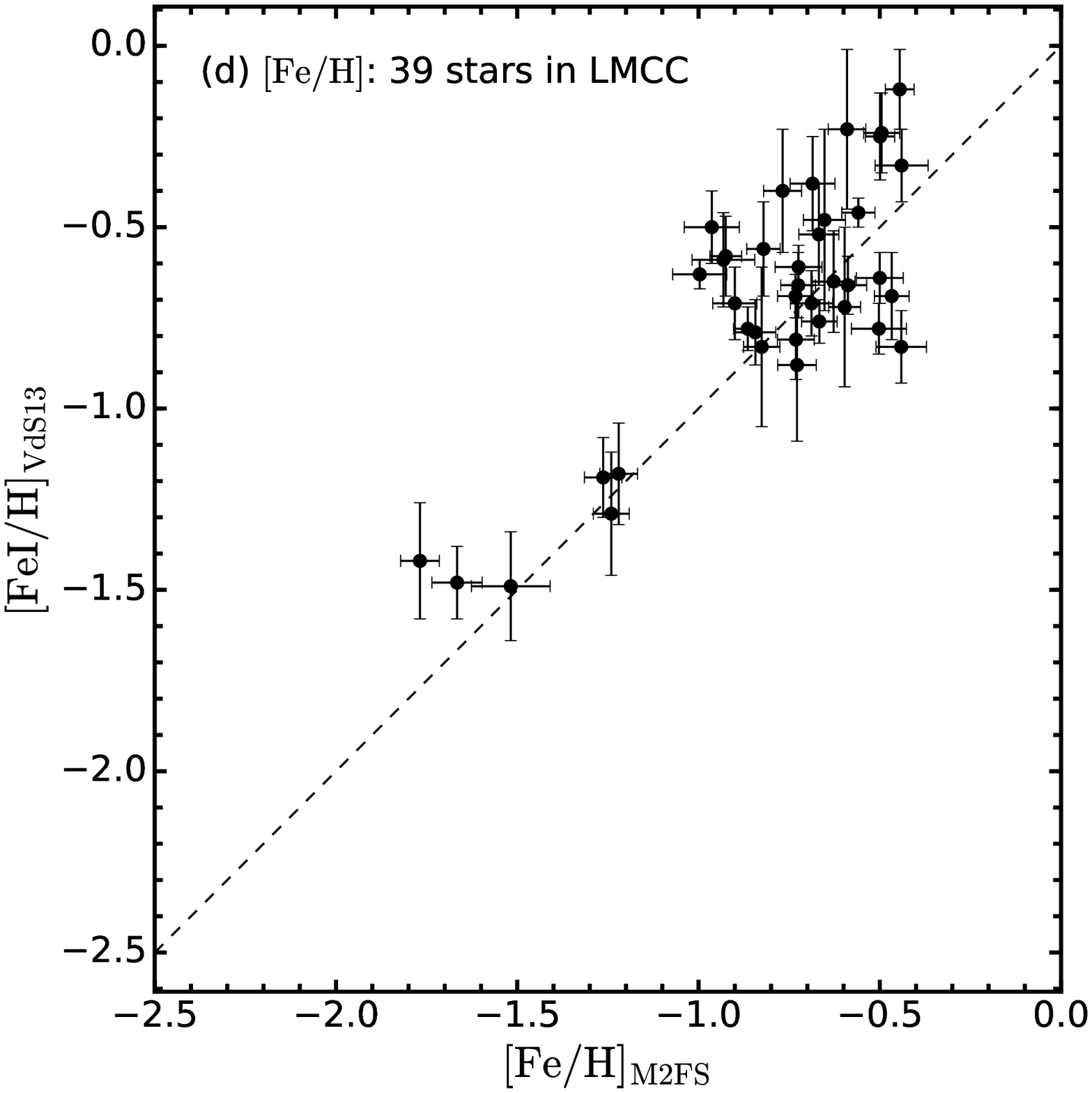}
      \includegraphics[width=0.4\textwidth]{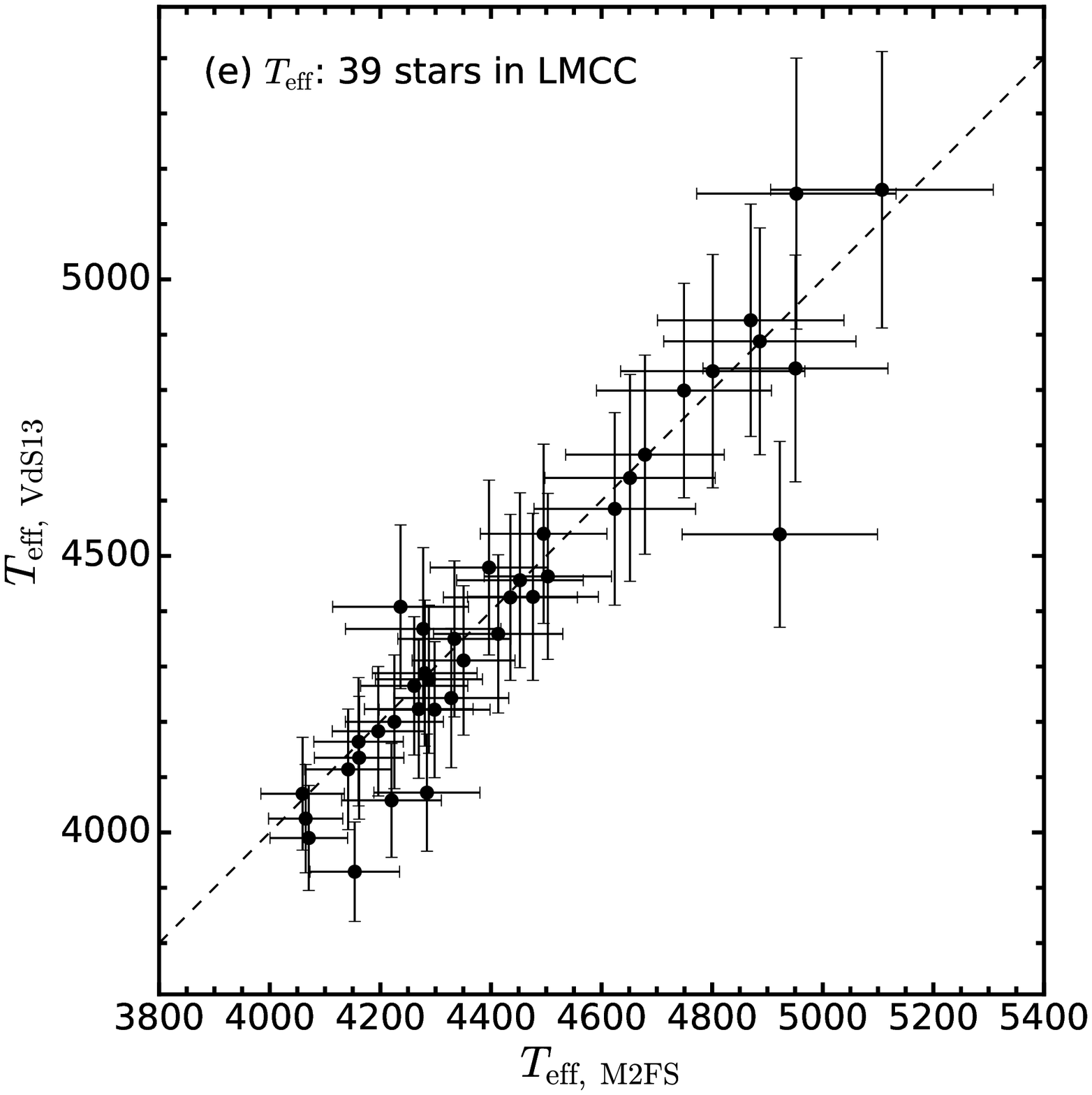}
      \includegraphics[width=0.4\textwidth]{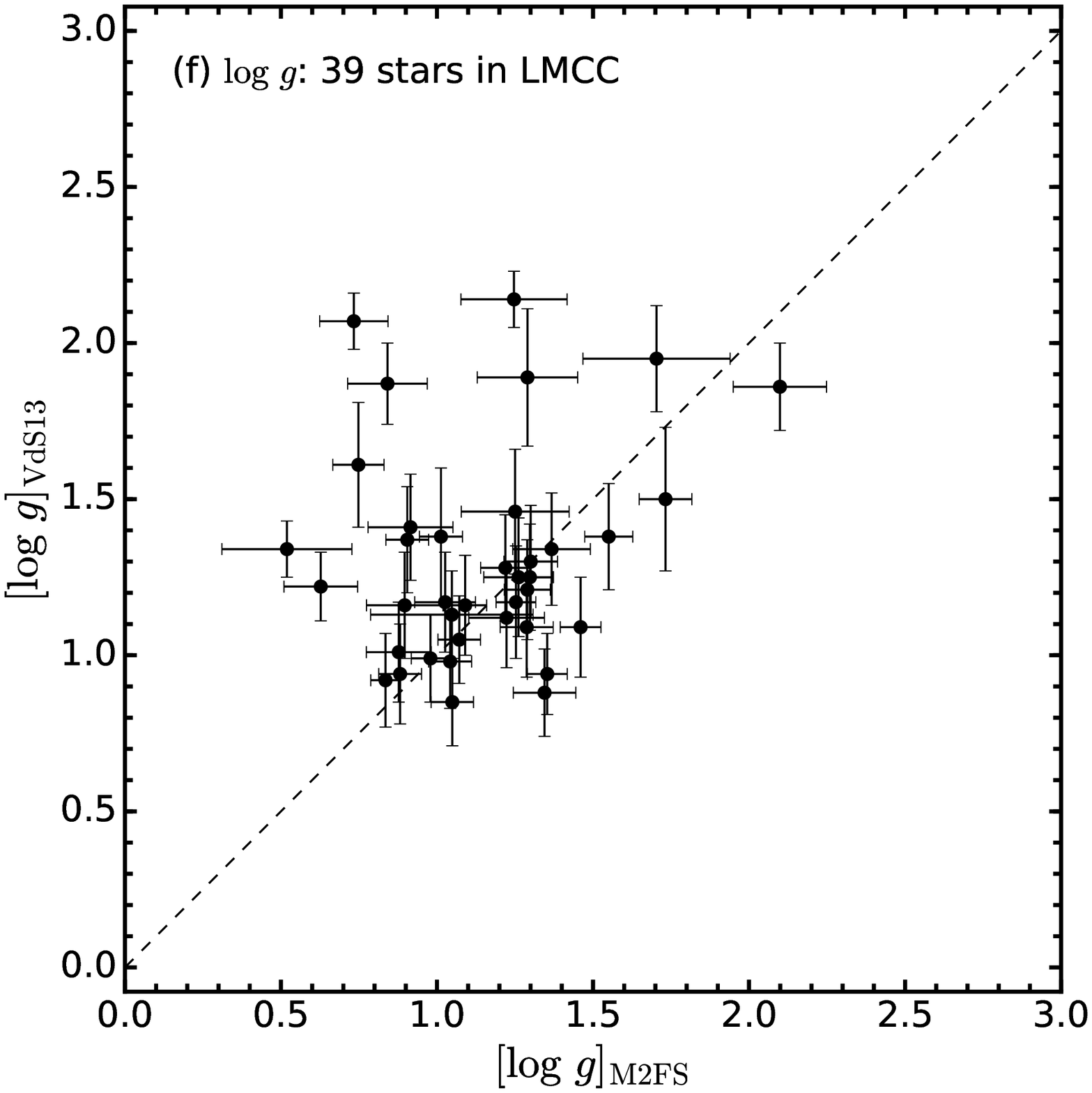}
   \caption{Comparison of the physical parameters measured in this study to those measured by C05 (top panels) and VdS13 (middle and bottom panels). Dashed lines indicate 1:1 relation.
   Note that the error bars of $v_{\rm los,\,M2FS}$ and $v_{\rm los,\,VdS13}$ in (a) and (c) are smaller than size of the dots.}
   \label{fig:comp}
\end{figure*}

\begin{figure*}[htbp]
   \centering
     \includegraphics[width=0.45\textwidth]{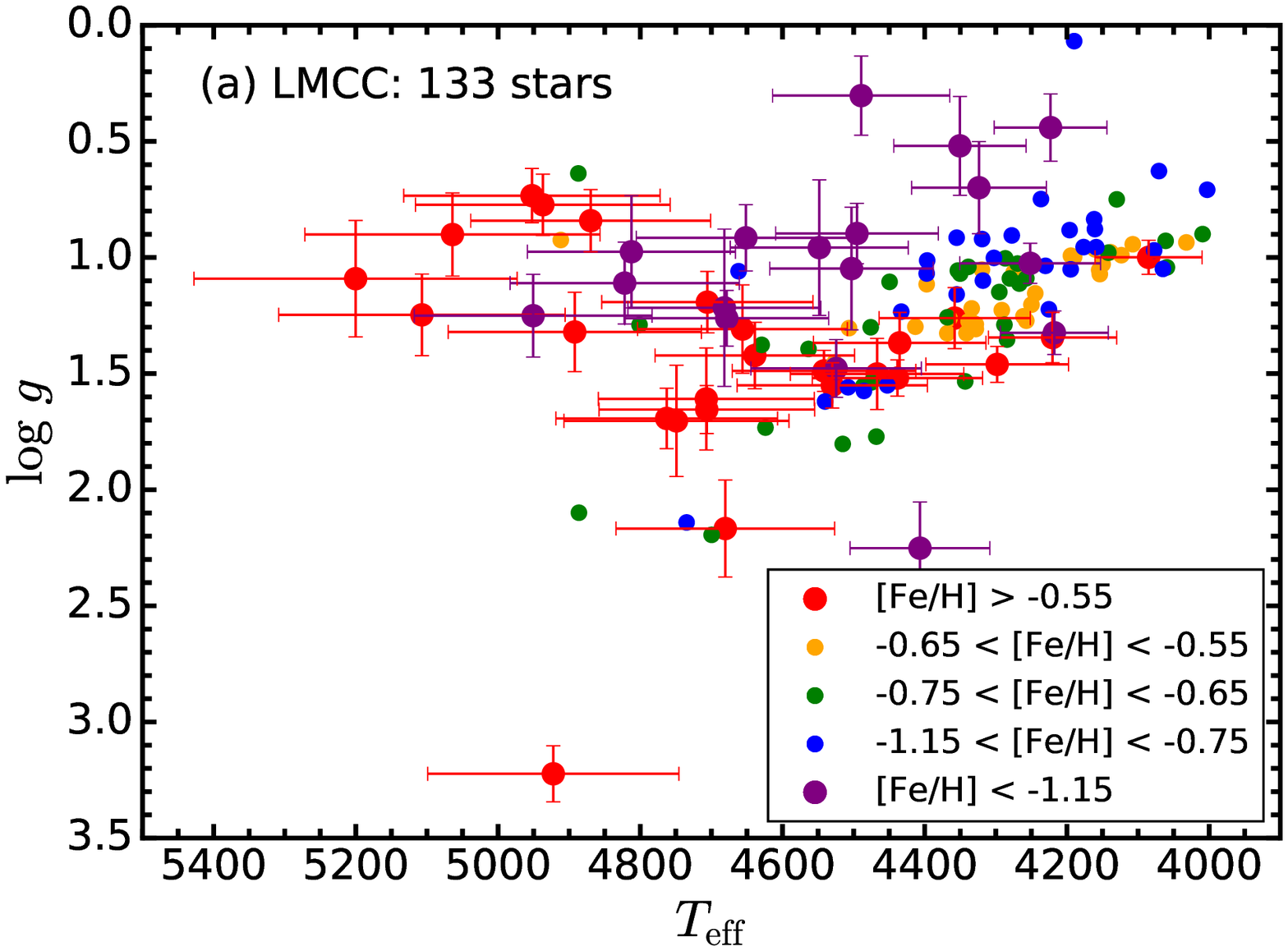} 
     \includegraphics[width=0.45\textwidth]{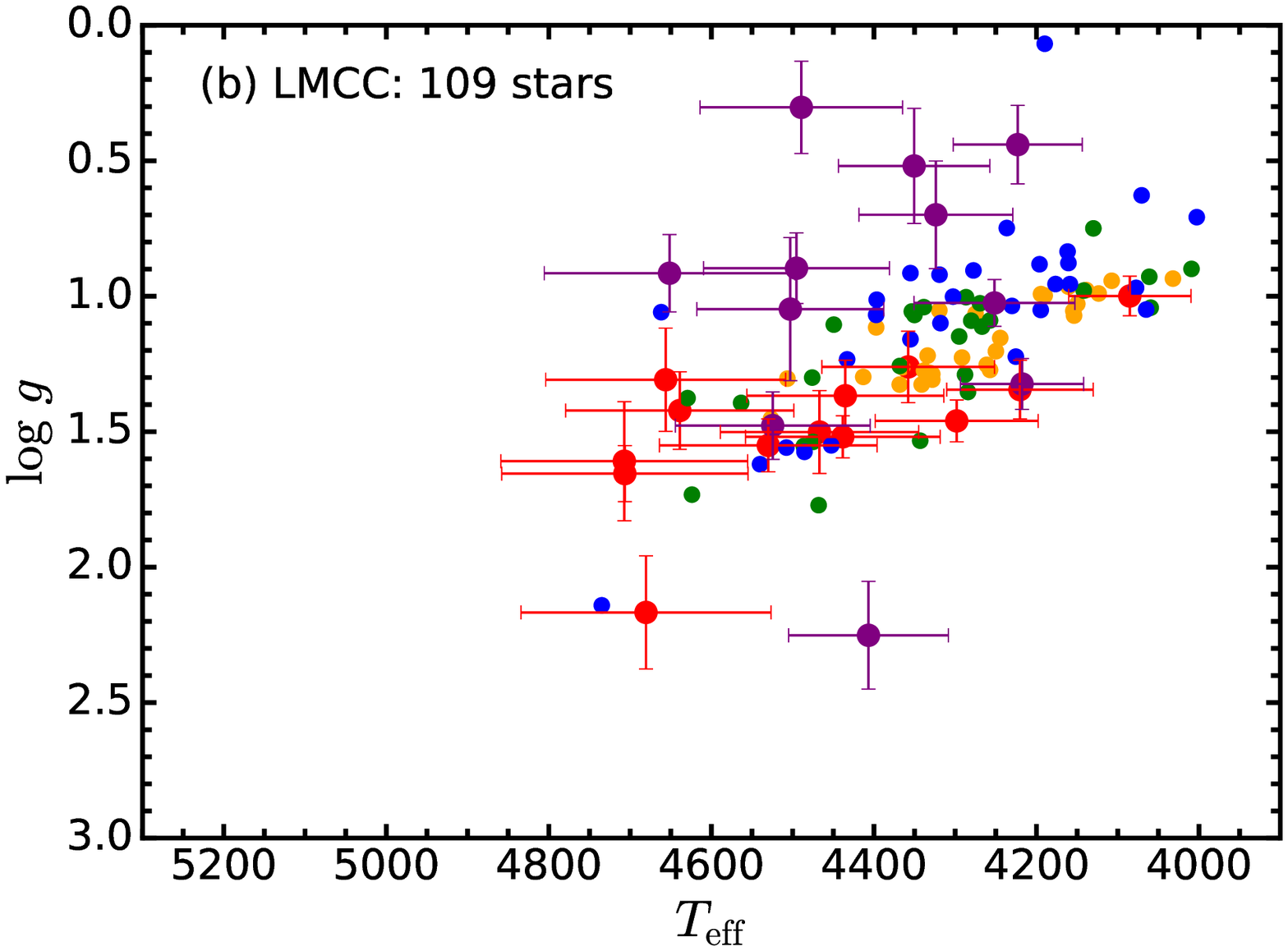} 
     \includegraphics[width=0.45\textwidth]{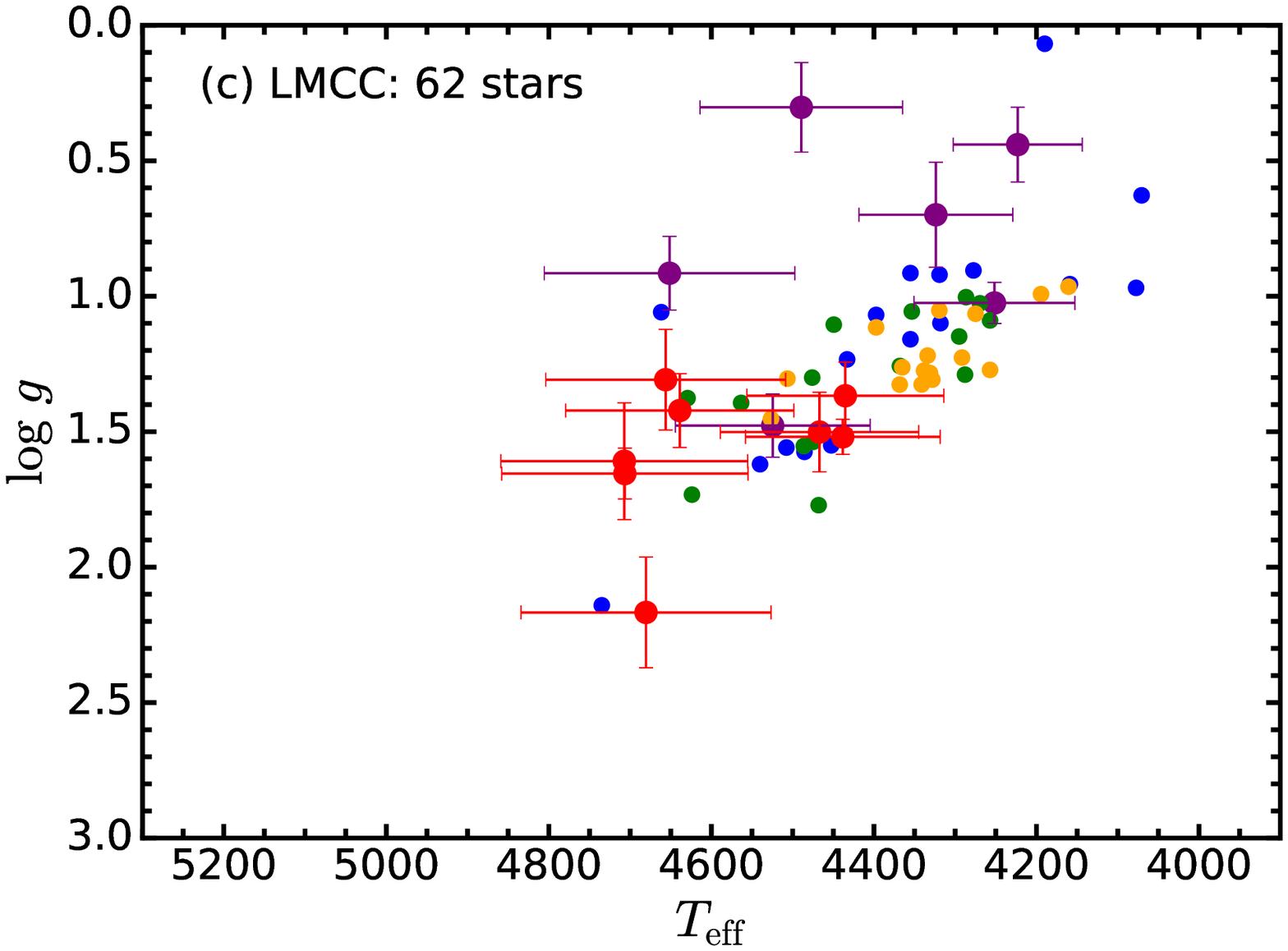} 
     \includegraphics[width=0.45\textwidth]{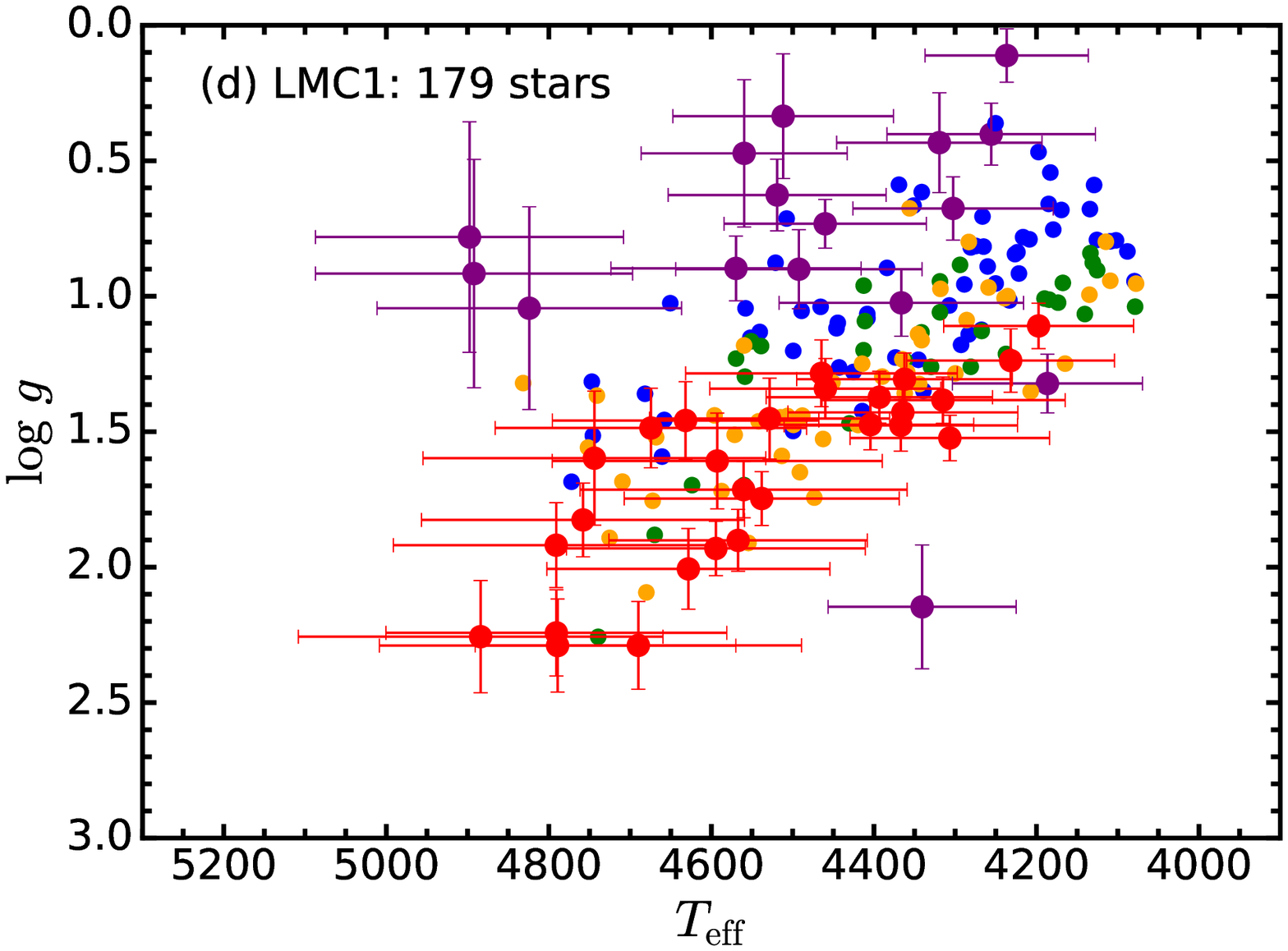} 
   \caption{The spectroscopic HR diagram on the $T_{\rm eff}$-$\log{g}$ plane color-coded by metallicity bins (see the legend). Note that we only plot the error bars of the stars in the most metal-rich (red) and the most metal-poor (purple) bins.}
   \label{fig:HRD}
\end{figure*}

\begin{figure*}[htbp]
   \centering
   \includegraphics[width=0.49\textwidth]{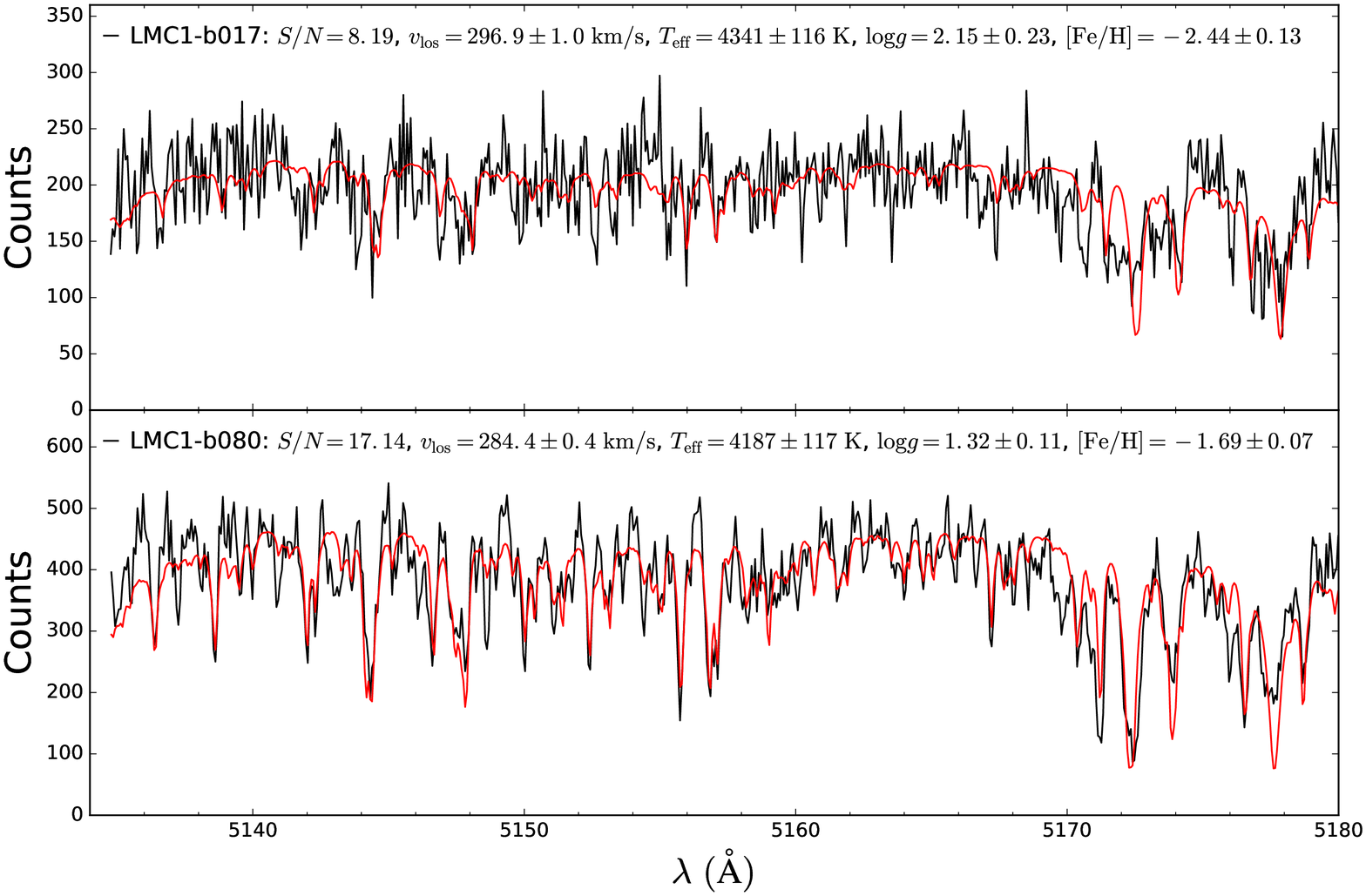}
   \includegraphics[width=0.49\textwidth]{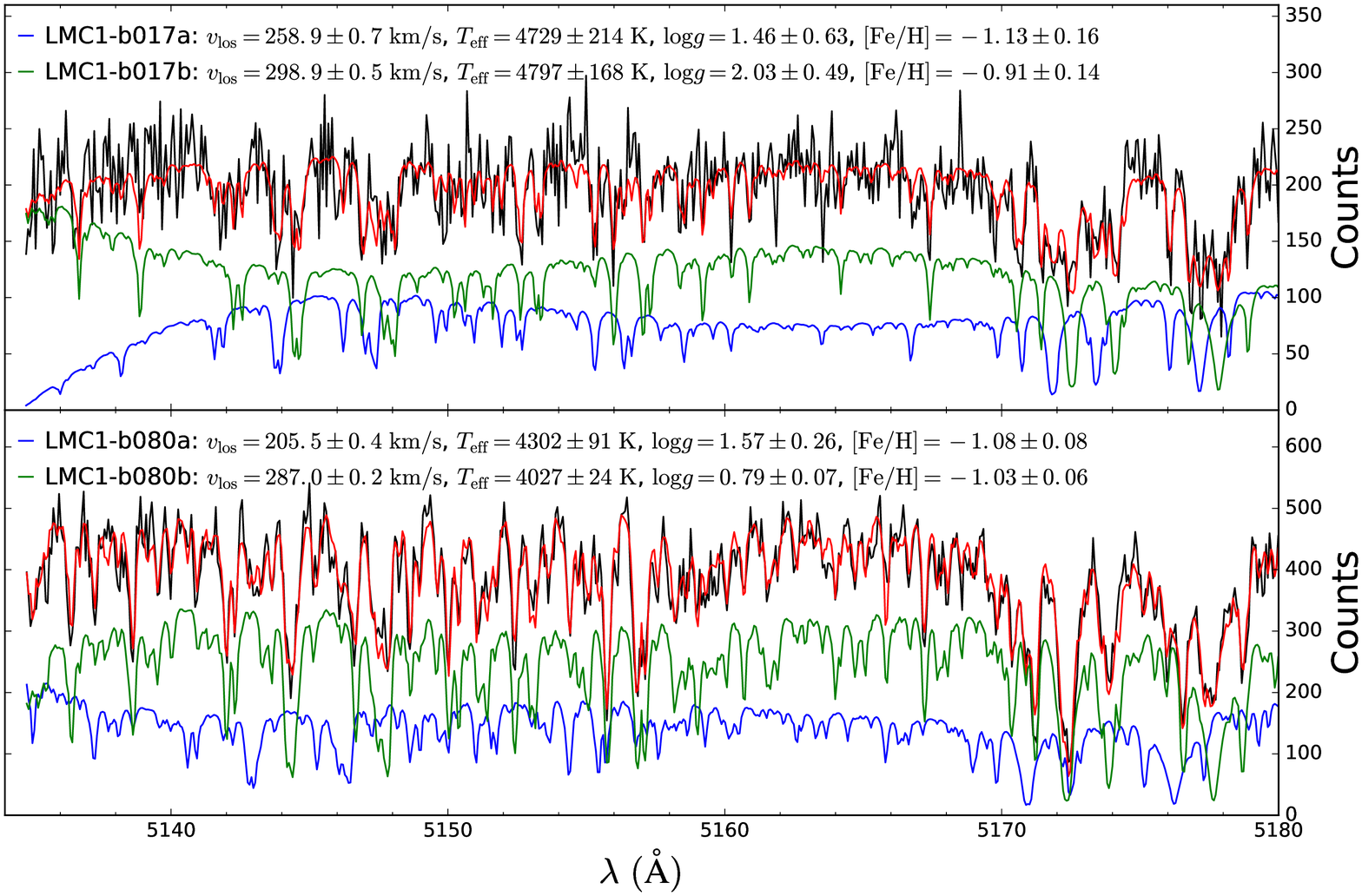}
   \caption{The spectra (black) and their best-fit models (red) for the two anomalous stellar targets in the LMC1 sample. Left panels: results of a single-star model; right panels: results of a double-star model with the first component in blue and the second one in green. The best-fit parameters are also shown in each panel. }
   \label{fig:spec_double}
\end{figure*}

One problem with all the MDFs mentioned above is that they may reflect
biases due to the sample selection effects.  Since the LMC1 sample was
selected in a very specific manner (see Section \ref{sec:selection} and
Figure \ref{fig:CMD_selection}), it is feasible for us to correct the
LMC1 MDF for its underlying selection bias.  Our correction starts by
counting stars within every rectangle shown in
Figure \ref{fig:CMD_selection}.  Within a given rectangle, the sub-MDF is
corrected by multiplying a weight factor, which is a ratio of the
total number of stars divided by the observed number of stars.  This
step is then carried out for all the rectangles, and all sub-MDFs are
summed up before being renormalized to a final corrected MDF.  The
comparison between the corrected MDF and the `raw' MDF from the
reduced LMC1 sample is shown in Figure \ref{fig:MDF}d.  The corrected LMC1 MDF
has a slightly higher mean for the metal-rich component but with the
same fraction, and can now be fitted as
$(w_1,\,\mu_1,\,\sigma_1)=(0.85\pm0.08,\,-0.66\pm0.02,\,0.17\pm0.01)$
and
$(w_2,\,\mu_2,\,\sigma_2)=(0.15\pm0.08,\,-1.20\pm0.24,\,0.41\pm0.06)$.
This correction can be applied in a reproducible manner to the LMC1
sample, but not to the LMCC sample due to the different and
ill-defined selection method adopted by C05.  As a result, the
corrected MDF based on 177 stars in the LMC1 sample represents our
best estimate of the MDF of the central LMC bar.

\subsection{Kinematics}

The heliocentric LOS velocities range between 175.2~km~s$^{-1}$ and
343.1~km~s$^{-1}$ for our full RG sample (including all the LMCC and
LMC1 RGs), implying that all stars in our sample should be considered
as LMC members according to \citet{Zhao03}.  The mean heliocentric LOS
velocity of 131 LMCC stars is $262.9\pm2.1$~km~s$^{-1}$ with a standard deviation of
$24.4\pm1.6$~km~s$^{-1}$, compared to $258.1\pm2.1$~km~s$^{-1}$
and $28.3\pm1.5$~km~s$^{-1}$ for 177 LMC1 stars.  For a combined
sample of 239 RGs (i.e. all stars in Figures 6c and 6d), we measured a
mean heliocentric LOS velocity of $258.5\pm1.8$~km~s$^{-1}$ with a
velocity dispersion of $27.8\pm1.3$~km~s$^{-1}$, in good agreement
with previous measurement of the LMC bar by C05.

In Figure \ref{fig:kinematics}, we plot the heliocentric LOS velocities (left panels) and the corresponding velocity dispersions (right panels) both as a function of metallicity. 
The top panels show the kinematics of LMCC and LMC1 separately, while the bottom panels show the combined results from the two fields. 
In all cases, the whole sample is divided into four subsamples according to the metallicity bins in Table \ref{tab:kinematics}, in which we also report the mean heliocentric velocity and the corresponding velocity dispersion in each bin based on the sample of 239 RGs.
It can be seen both in Figure \ref{fig:kinematics} and Table \ref{tab:kinematics} that the velocity dispersion increases significantly from $18.7\pm1.9$~km~s$^{-1}$ for the most metal-rich stars to $31.2\pm4.3$~km~s$^{-1}$ for the most metal-poor stars. 

\section{Summary and Discussion}
\label{sec:discussion}
In this paper, we have measured physical parameters for 312
stellar targets in the LMC bar from high-resolution spectra obtained
with the multi-fiber facility M2FS on the \textit{Magellan}/Clay
Telescope.  Assuming $E(V-I)=0.15\pm0.07$ mag constantly, we initially
estimated an effective temperature of each star from the OGLE-II $V-I$
color \citep{Udalski00} according to the color-temperature relation
for giants \citep{RM05b}.  Then, each star's LOS velocity, surface
gravity and metallicity were fitted simultaneously by comparing
each background-subtracted spectrum statistically with a library of
template spectra made by synthetic modeling \citep{Lee08a, Lee08b}.

All of our stellar spectra are obtained from two LMC bar fields,
labelled as `LMCC' and `LMC1', respectively.  Four of the 312 stellar
targets are confirmed to be possible double-star sources. Among the
remaining 308 stars, 177 LMC1 RGs are observed spectroscopically for
the first time. The remaining 131 LMCC RGs were first observed by
C05 and 39 were also subsequently observed by
VdS13.  The reobserved stars are used to directly compare
the measurements of the physical parameters among studies.
As a
result, we found that our metallicity measurements gave a MDF with
lower peaks of both metal-rich and metal-poor populations, comparing
to C05.  The differences in these populations are $\rm
\Delta[Fe/H]=0.29\pm0.02~dex\ and\ 0.12\pm0.29~dex$, respectively, given the typical
measurement errors in metallicity are 0.06~dex and 0.12~dex for us and
C05.

Based on 177 RGs in the LMC1 sample, we measured a new mean
metallicity in the LMC bar of $\rm [Fe/H]=-0.76\pm0.02$~dex ($\sigma=0.28\pm0.03$~dex) and generated a
new sample-selection-effect corrected MDF for the LMC bar
(Figure \ref{fig:MDF}d).  The MDF can be best fitted by two Gaussian
distributions with a portion of $85\%$ and $15\%$, respectively.  The
majority population is made of the more metal-rich stars and has a
mean value of $\rm [Fe/H]=-0.66\pm0.02$ dex ($\sigma=0.17\pm0.01$
dex), while the minority one is relatively metal-poor and has a mean
$\rm [Fe/H]=-1.20\pm0.24$ dex ($\sigma=0.41\pm0.06$ dex).  Our
newly-observed MDF is different from that reported by
C05 in two aspects: first, our metal-rich population
peaks of about $0.29\pm0.02$ dex lower than that found by
C05; second, our metal-poor population fraction
($\approx 15$ \%) is slightly larger than that ($\approx 11$ \%) found
by C05.

In both LMCC and LMC1 samples, the kinematics as a function of
metallicity provides a good way to trace the evolution of the LMC bar
(Figure \ref{fig:kinematics}).  The kinematics also show a clear trend of
the decreased velocity dispersion with increasing metallicity.  This
trend not only confirms that more metal-poor stellar populations are
distinguished dynamically by metallicity, but also implies a true
chemo-dynamical evolution of the bar that covers at least a few
crossing times of the system.  C05 found that there
may exist an old, thicker disk or halo population in the LMC bar with
a velocity dispersion $40.8\pm 1.7$ km~s$^{-1}$ among the most
metal-poor stars in their sample ($5\ \%$).  Our results indicate the
existence of the thicker component in the LMC bar, though a smaller
value $31.2\pm4.3$ km~s$^{-1}$ is measured from our most metal-poor
($\approx10$ \%) stars.  This reflects the kinematics of the old
populations in the LMC bar, and is in good agreement with the velocity
dispersion of 30 km~s$^{-1}$ for RR Lyrae stars \citep{Gratton04}.

\subsection{Stellar Masses:  A Cautionary Tale}
\label{sec:mass}
\begin{figure*}[htbp]
   \centering
   \includegraphics[width=0.75\textwidth]{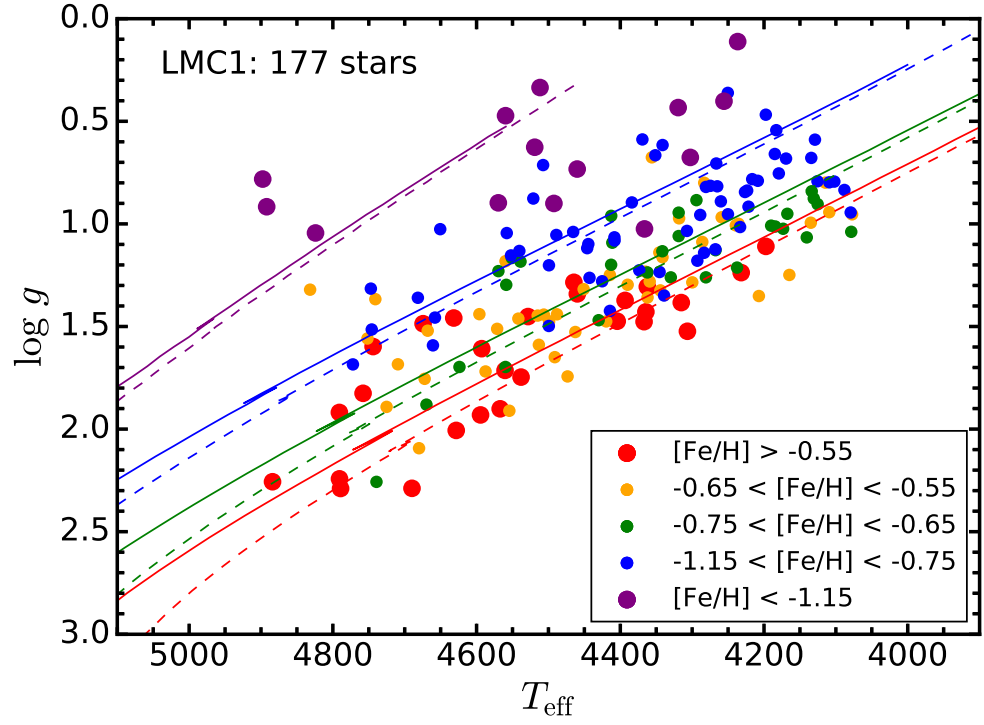}
   \caption{The HR diagram similar to Figure \ref{fig:HRD}d but with comparison to PARSEC isochrones \citep{Bressan12} of 5 Gyr (solid curves) and 12.7 Gyr (dashed curves), respectively. The color dots are coded by metallicity as shown in the legend. The isochrones are color-coded by $\rm [Fe/H]=-0.5$ (red), $-0.7$ (green), $-1.1$ (blue) and $-2.2$ (purple), respectively.  We applied a systematic offset of 0.3 dex in \logg\ of the isochrones. }
   \label{fig:HRD_iso}
\end{figure*}

\begin{figure*}[htbp]
   \centering
  \includegraphics[width=0.45\textwidth]{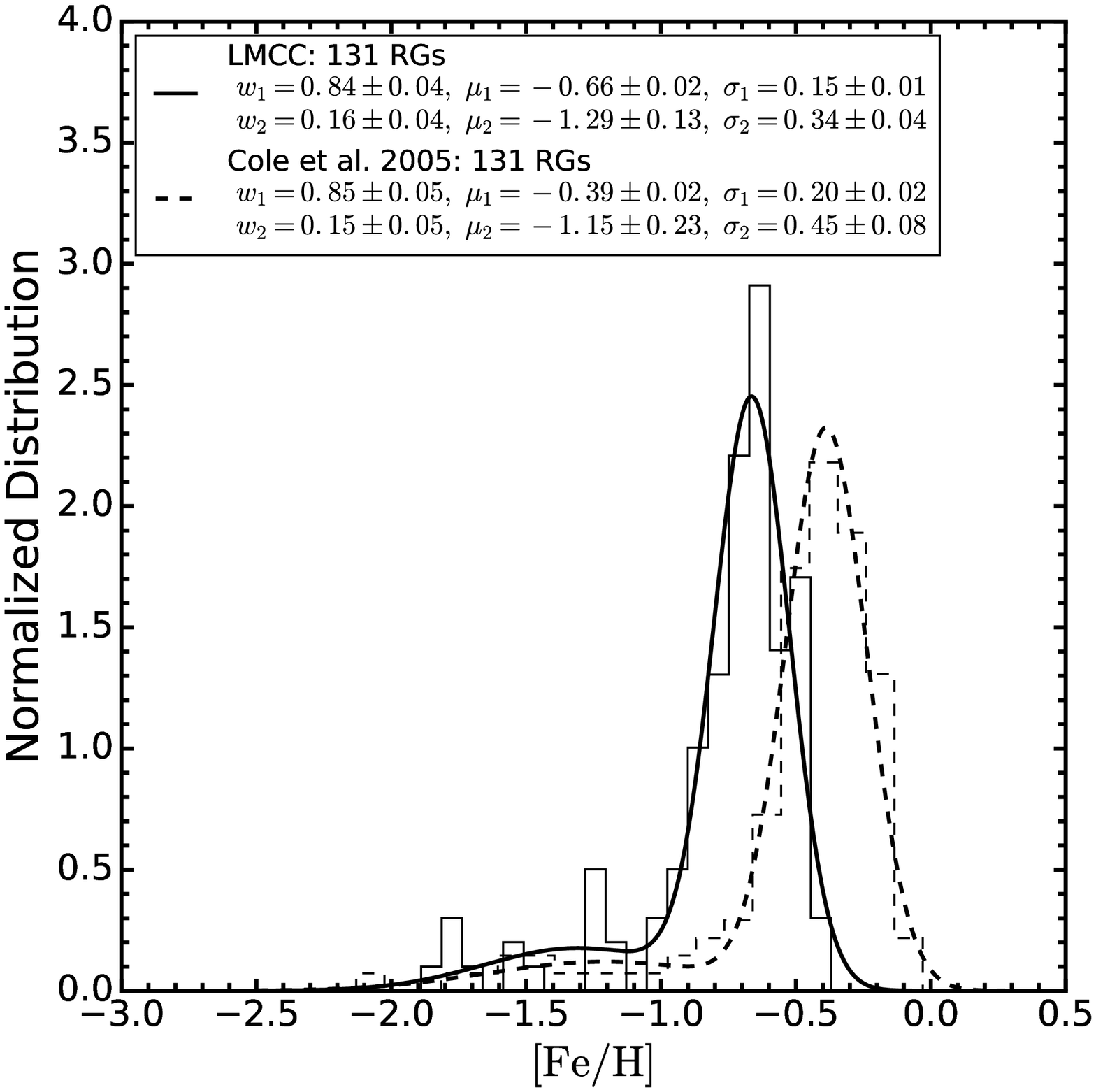}
  \includegraphics[width=0.45\textwidth]{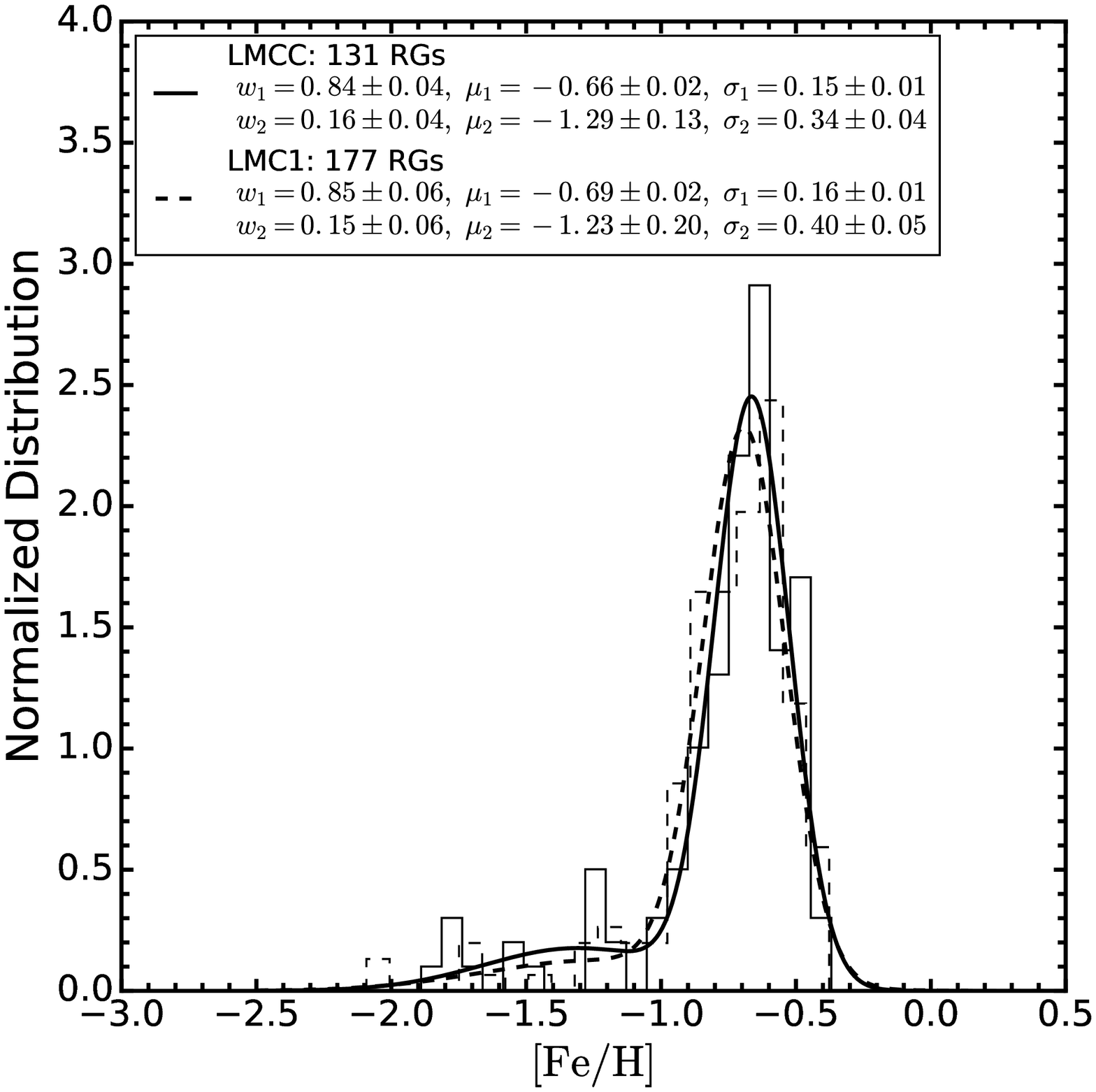}
  \includegraphics[width=0.45\textwidth]{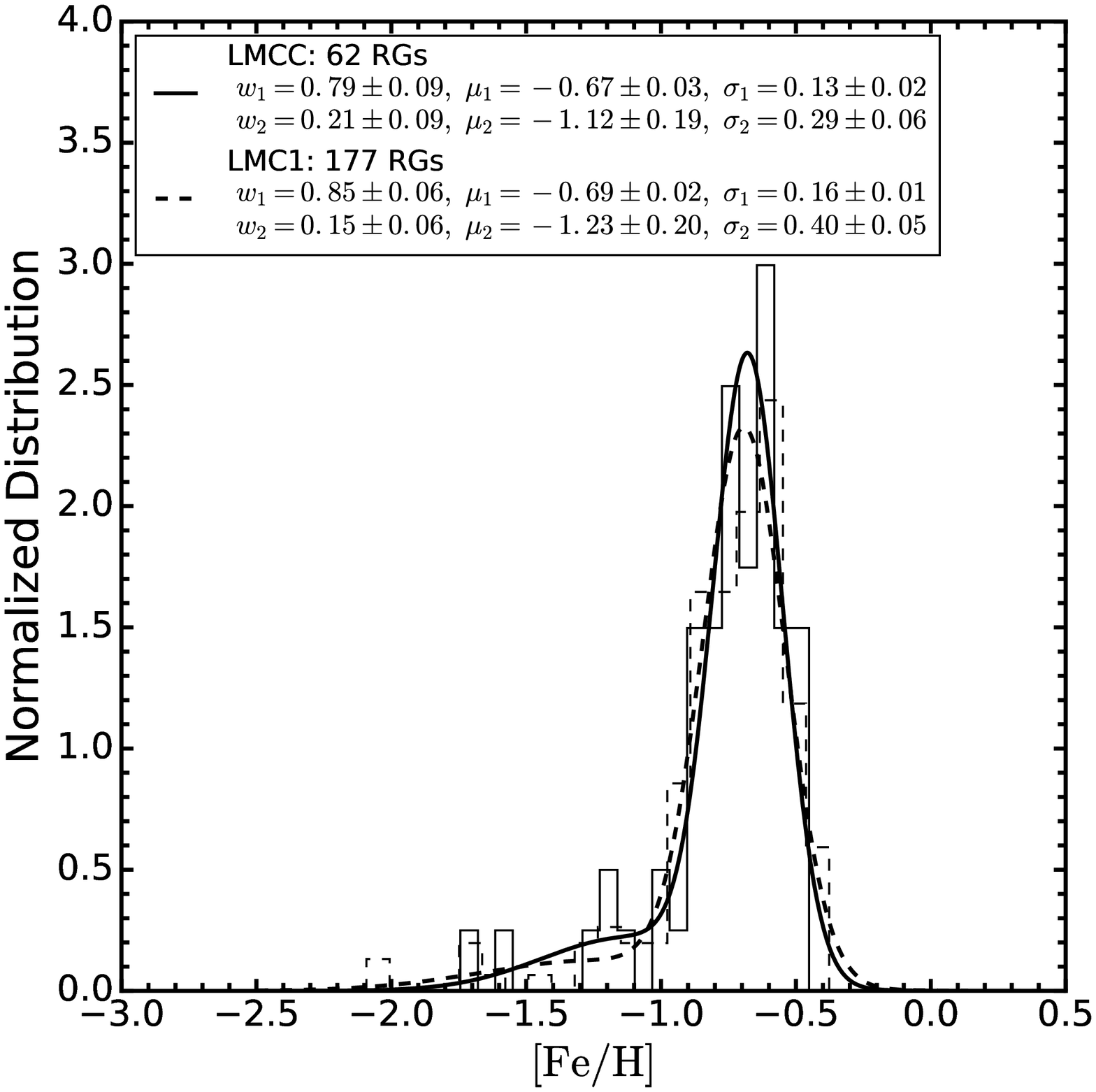}
  \includegraphics[width=0.45\textwidth]{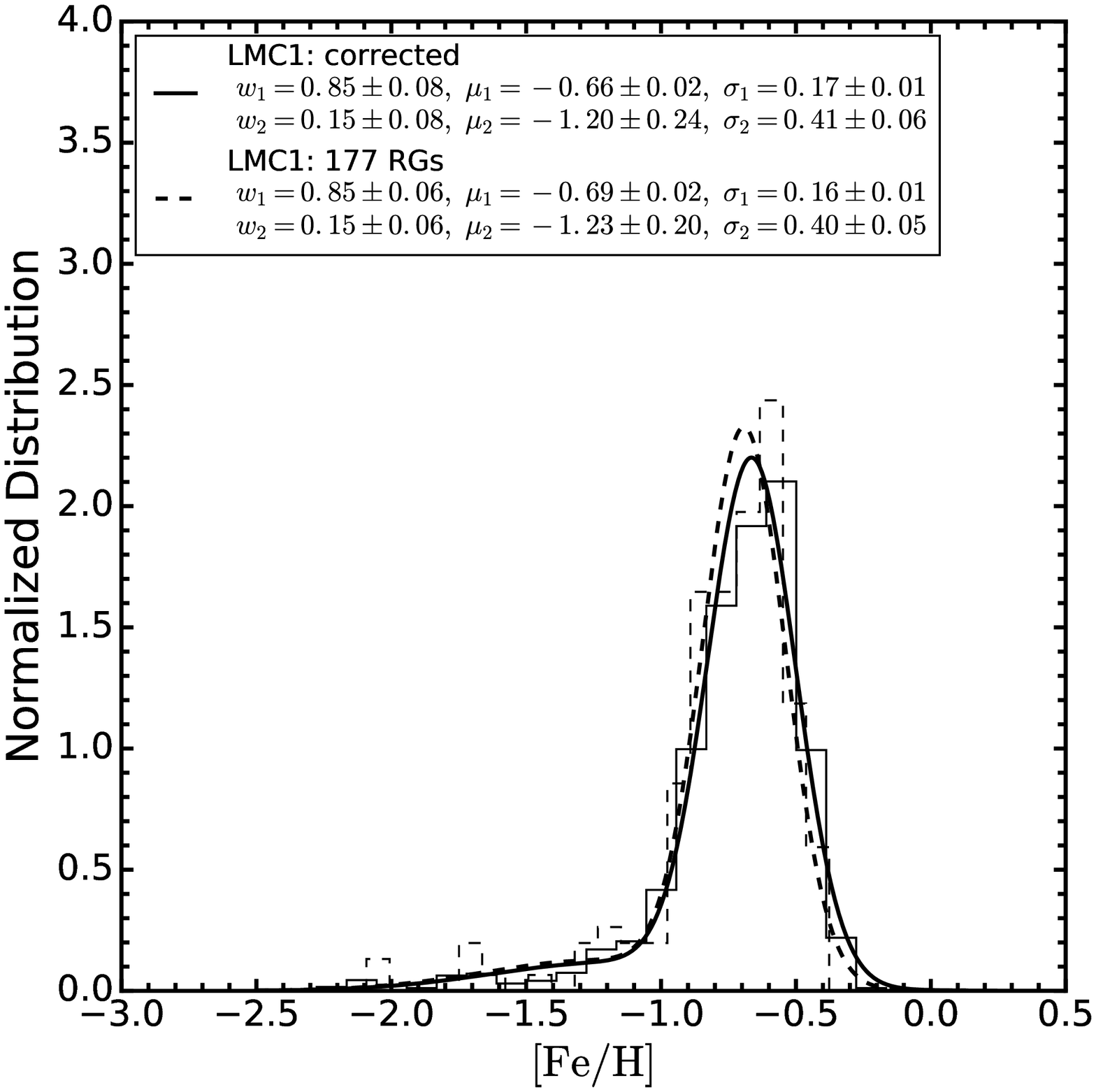}
   \caption{
   The MDFs from different RG samples with the best fits by a mixture of two Gaussian distributions.
   The legends list the normalized weight $w_i$, mean $\mu_i$ and standard deviation $\sigma_i$ for each Gaussian component. 
   (a) Top-left panel shows the MDFs of the same 131 RGs with metallicities measured with M2FS (solid) versus that measured by C05 (dashed). 
   (b) Top-right panel shows the MDFs of 131 RGs from LMCC (solid) and of 177 RGs from LMC1 (dashed), using the metallicities measured with M2FS only. 
   (c) Bottom-left panel shows the MDF of 62 RGs from LMCC (solid) that locate in the same LMC1 sample region on CMD, using the metallicities measured with M2FS only; the dashed MDF is the same as that in the top-right panel. 
   (d) Bottom-right panel: the MDF of 177 LMC1 RGs with selection effect correction (solid), comparing to the `raw' MDF (dashed).
 }
   \label{fig:MDF}
\end{figure*}

\begin{figure*}[htbp]
   \centering
     \includegraphics[width=0.45\textwidth]{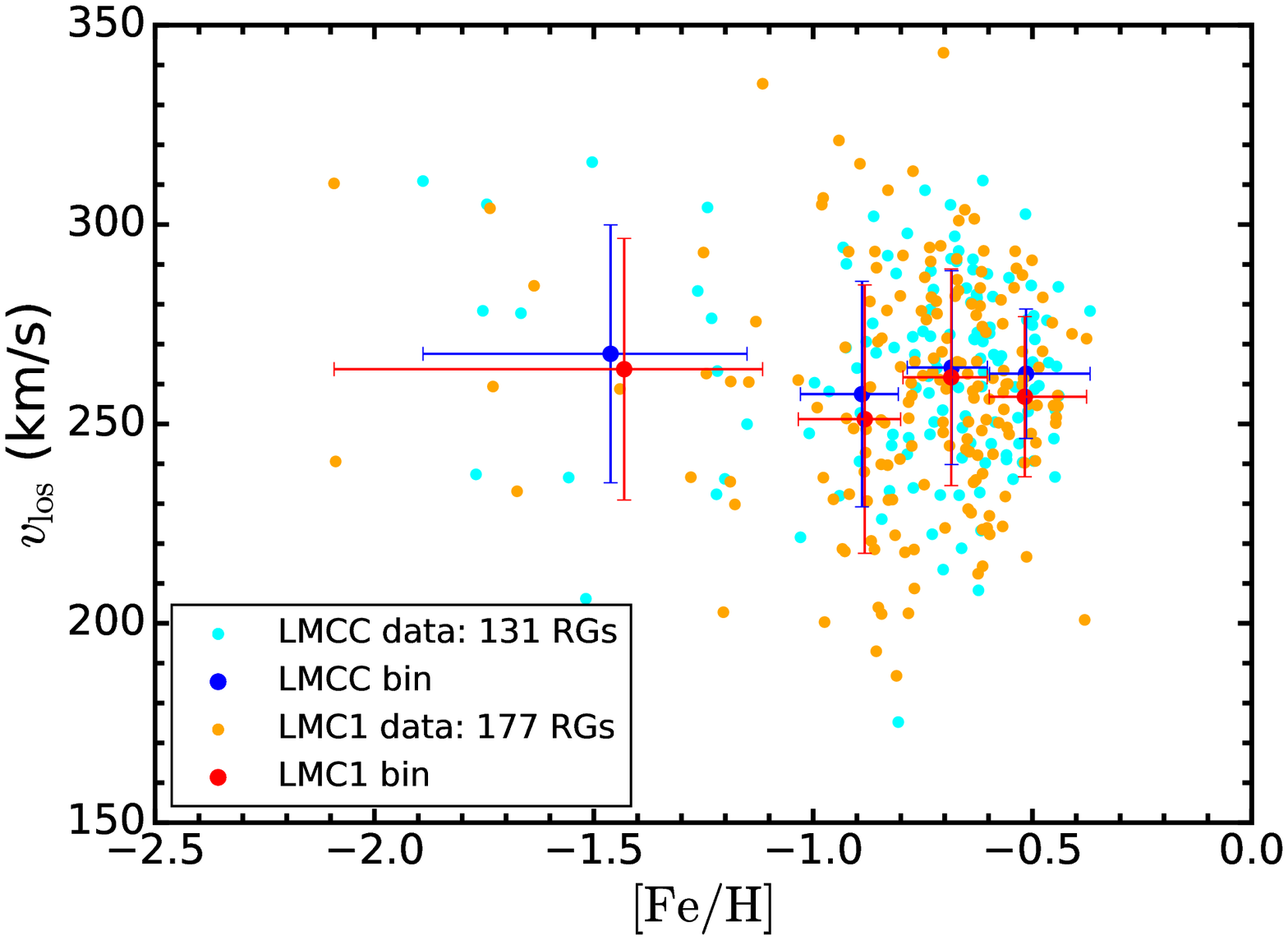}
     \includegraphics[width=0.45\textwidth]{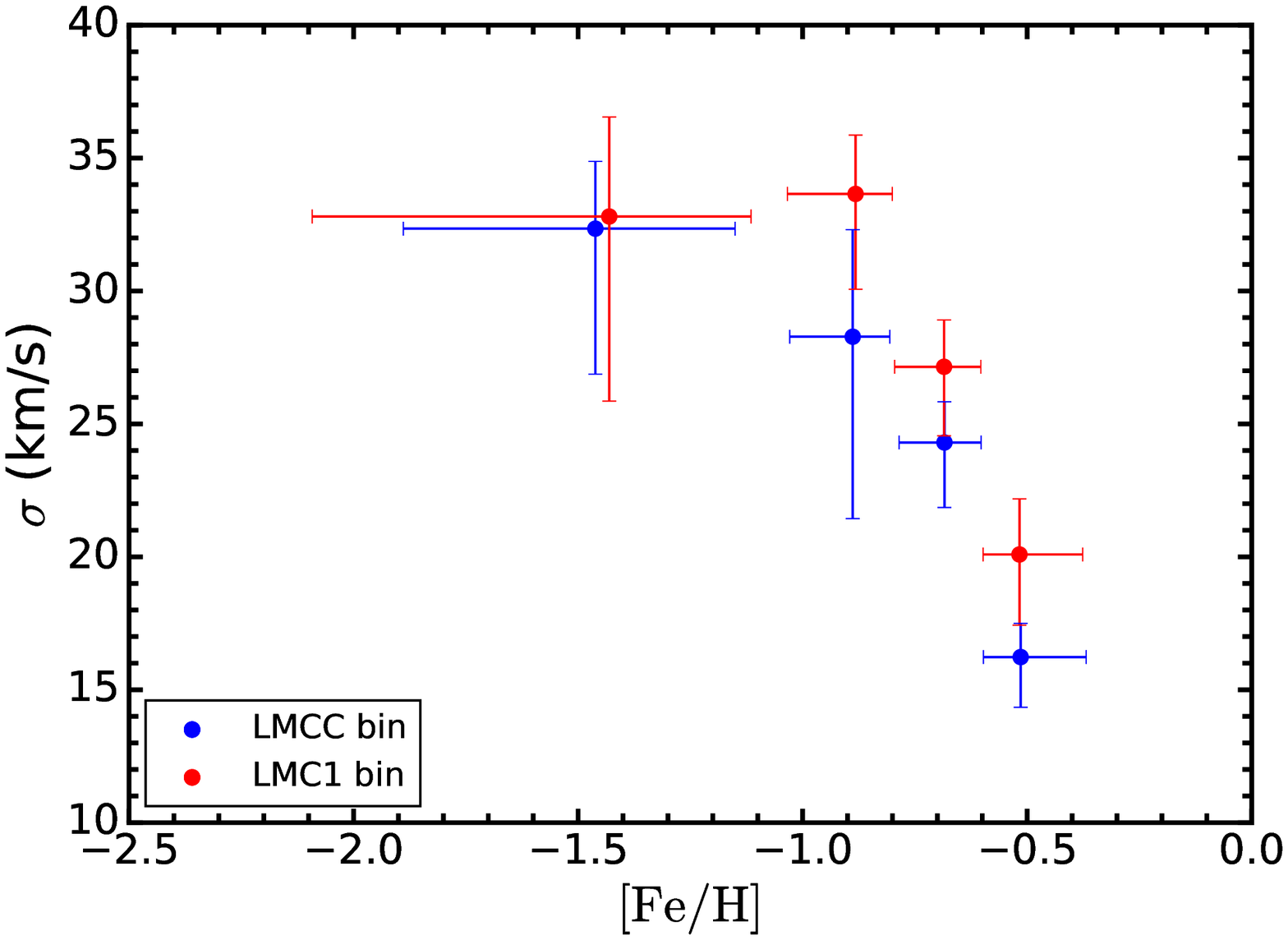}
     \includegraphics[width=0.45\textwidth]{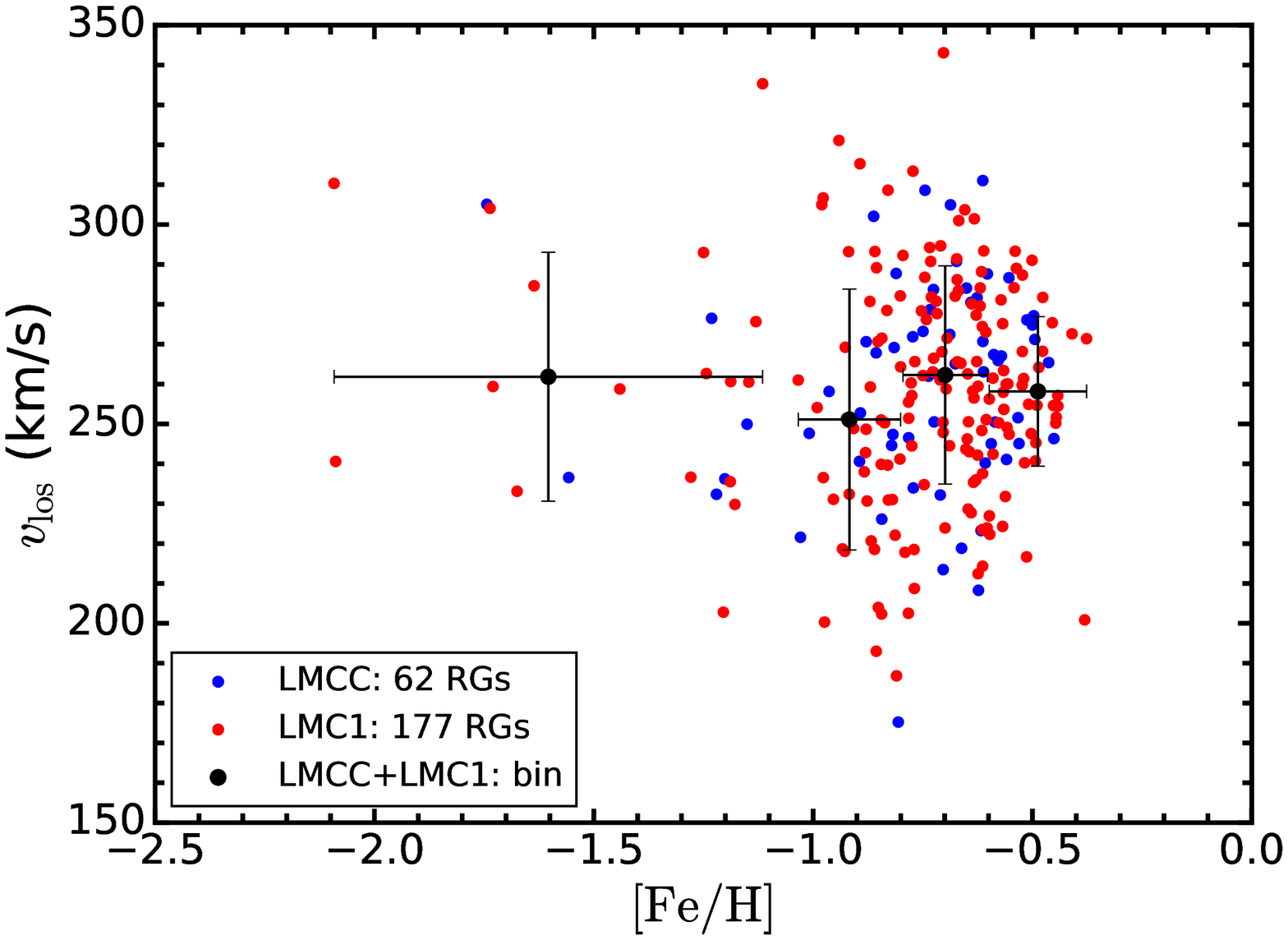}
     \includegraphics[width=0.45\textwidth]{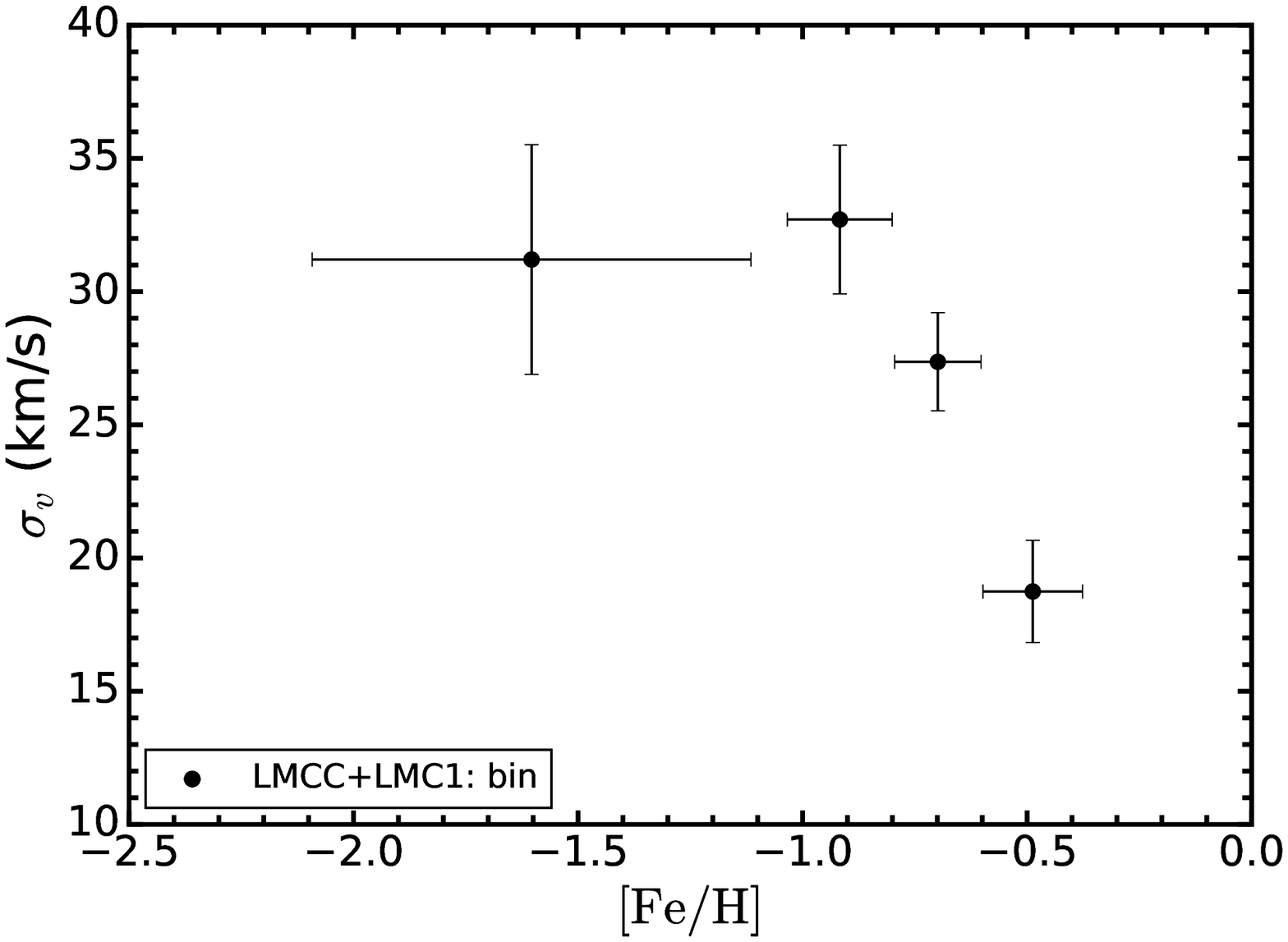}
   \caption{The heliocentric LOS velocities (left panels) and velocity dispersions (right panels) from M2FS spectra as a function of metallicity. The samples are binned by metallicity as shown in Table \ref{tab:kinematics}. The top panels show the LMCC and LMC1 samples separately, and the bottom panels show their combined sample. The horizontal error bars indicate the full range values in each bin, and the vertical error bars are the standard deviations of the velocities (left panels) and velocity dispersions (right panels) in each bin, respectively.}
   \label{fig:kinematics}
\end{figure*}

The ultimate aim of this study is to map the chemo-dynamical
properties as a function of position and age, such as the
age-metallicity relation and the star formation history of the LMC
bar.  To take the first step, we compared our spectroscopic results
with PARSEC isochrones \citep{Bressan12} in Figure \ref{fig:HRD_iso},
which provides a way to measure the age of a RG if its mass is well
constrained.

\begin{figure*}[htbp]
   \centering
   \includegraphics[width=0.75\textwidth]{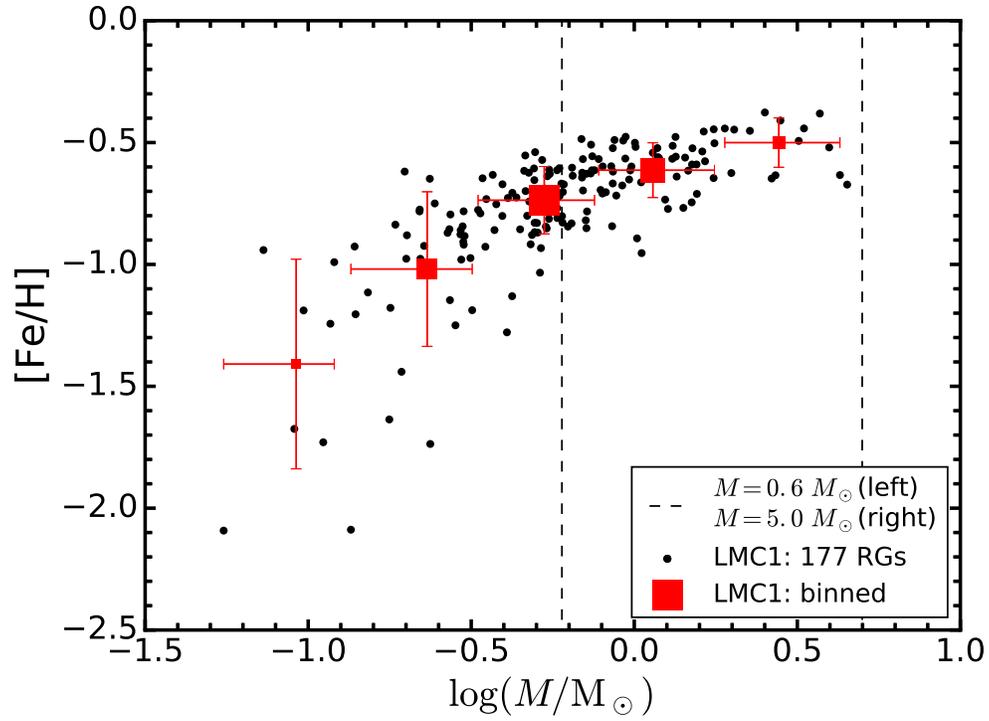}
   \caption{The masses of the LMC1 RGs as a function of metallicity. The black dots represent the individual stars. The red squares represent the binned results (mean values in $\rm [Fe/H]$ and $\log{M}$), with their sizes indicating the sample sizes; the horizontal error bars are made to cover the full mass ranges in each bin, while the vertical error bars show the standard deviations of metallicity.}
   \label{fig:MMR}
\end{figure*}

To calculate the masses of the LMC1 RGs, we followed the method
introduced by \citet{Feuillet16} who used the well-known relations
$L=4\pi R^2 \sigma T^4_{\rm eff}$ and $g=GM/R^2$ (where $\sigma$ is
the Stefan-Boltzmann constant and $G$ is the gravitational constant).
In our calculation, the luminosities were calculated from apparent
OGLE-II $V$-band magnitudes, the distance modulus to the LMC
\citep[$18.50\pm0.06$~mag, ][]{pietrzynski09}, a constant extinction
($A_V=0.34$~mag), and the bolometric corrections (BCs) interpolated
from the model-based BC tables \citetext{the dustyAGB07 database,
  \citealp{Marigo08, Girardi08}}.  We also used the solar bolometric
magnitude $M_{\rm Bol,\,\odot}=4.77$~mag and luminosity
$L_{\odot}=3.844\times 10^{33}$~erg/s \citep{Bahcall95}.  Shown in
Figure \ref{fig:MMR} are the calculated masses as a function of
metallicity.  To elucidate the relation between mass and metallicity,
we binned our sample by mass and calculated a mean metallicity for
each bin (red squares).  It can be seen that the mean metallicity
increases with mass increasing, or equivalently with age decreasing.
This phenomenon is similar to the one found by C05 on
their age-metallicity relation.

But an important problem with our mass calculation is evident in
Figure \ref{fig:MMR}: The expected mass of RGs in this portion of the CMD
would be expected to be in a range spanning approximately from 0.6 to
5~$\rm M_\odot$ (see the two dashed vertical lines in Figure \ref{fig:MMR}).
However, a significant number of our RGs have calculated masses below
the lower limit.  Due to this anomalously large range of masses from
our simple calculation, it seems impossible to derive reliable ages
from any RG mass-age relation (e.g. from the PARSEC isochrones).  The
origin of the discrepancy in our mass/age scale remains unclear and
would not be solved by a simple offset in $\log{g}$ or in our $T_{\rm
  eff}$ scale.  We note that C05 and
\citet{Feuillet16} encountered similar problems, though to a different
quantitative extent.  Therefore, we limit ourselves in concluding that
there is an age-metallicity relation within the bar with a positive
slope over most of the age sampled by normal RGs.  Although the
potential for extracting the age-metallicity distribution from
multiplexed spectroscopy of field stars in the LMC bar is evident from
this discussion, improved stellar parameters---particularly
\logg\ values---will be needed to fully exploit this approach.

\acknowledgments We thank the anonymous referee for helpful
suggestions.  We thank Jeff Crane, Steve Shectman and Ian Thompson for
invaluable contributions to the design, construction and support of
M2FS.  M.M. and Y.-Y.S. are supported by National Science Foundation
(NSF) grant AST-1312997.  M.G.W. is supported by NSF grants
AST-1313045 and AST-1412999.  I.U.R. acknowledges support from the NSF
under Grant No. PHY-1430152 (JINA Center for the Evolution of the
Elements).

\bibliographystyle{apj}
\bibliography{main.bbl} 

\begin{table*}
\begin{center}
\caption{Summary of the observations in the LMC bar}
\begin{tabular}{cccccc}
\hline\hline
Field & $\alpha_{2000}$ & $\delta_{2000}$ & UT Date & Num. of Exp. & Total Exp. Time \\
Designation & (h:m:s) & ($^{\circ}$:$\arcmin$:$\arcsec$) & & & (s)    \\
\hline
LMC1 & 05:22:11.09 & -69:45:52.3 & Dec 23, 2014 & 3  & 3900 \\
LMC1 (Background) & &  & Nov 18, 2015 & 3 & 1800 \\
LMCC & 05:24:35.96 & -69:50:09.4 & Nov 15, 2015 & 3 & 5400 \\

\hline
\end{tabular}
\label{tab:obs}
\end{center}
\end{table*}

\begin{table*}
\begin{center}
\caption{Sample selected in the LMC bar}
\begin{tabular}{cccccc}
\hline\hline
ID & $\alpha_{2000}$ & $\delta_{2000}$ & V & I & Note$^a$ \\
 & (h:m:s) & ($^{\circ}$:$\arcmin$:$\arcsec$) & (mag) & (mag) &  \\
\hline
LMCC-b001 & 05:25:26.76 & -70:02:31.9 & $17.320\pm0.024$ & $15.893\pm0.024$ & 05252676-7002318 \\
LMCC-b002 & 05:25:25.99 & -70:01:43.0 & $17.353\pm0.015$ & $15.786\pm0.015$ & 05252599-7001429 \\
LMCC-b003 & 05:25:23.63 & -70:01:05.1 & $17.406\pm0.013$ & $16.012\pm0.013$ & 05252362-7001051 \\
LMCC-b004 & 05:25:31.26 & -70:01:02.9 & $17.418\pm0.013$ & $16.057\pm0.013$ & 05253125-7001028 \\
LMCC-b005 & 05:25:29.58 & -69:59:23.4 & $17.588\pm0.018$ & $16.496\pm0.018$ & 05252958-6959234 \\
 $\cdots$ & & & & & \\
LMC1-b002 & 05:23:17.12 & -69:55:29.9 & $17.654\pm0.064$ & $16.128\pm0.064$ & 5-138140 \\
LMC1-b003 & 05:23:14.46 & -69:53:45.1 & $17.983\pm0.049$ & $16.554\pm0.049$ & 5-145482 \\
LMC1-b004 & 05:23:37.49 & -69:52:42.3 & $17.590\pm0.035$ & $16.220\pm0.035$ & 5-145352 \\
LMC1-b005 & 05:23:21.80 & -69:51:35.1 & $17.967\pm0.053$ & $16.834\pm0.053$ & 5-153086 \\
LMC1-b006 & 05:23:32.77 & -69:50:51.2 & $18.036\pm0.069$ & $16.732\pm0.069$ & 5-153131 \\
 $\cdots$ & & & & & \\
\hline
\end{tabular}
\\
    \raggedright
    $^{a}$ Note: 2MASS ID for LMCC targets; OGLE-II ID for LMC1 targets\\
\label{tab:sample}
\end{center}
\end{table*}

\begin{table*}
  \scriptsize
  \centering
    \caption{Fits to twilight spectra for SSPP$^{a}$ library}
    \begin{tabular}{ccccccccc}
      \hline\hline
      \\
      Field & $\langle \overline{v_{\rm los}}-v_{\rm los,\odot}\rangle$ & 
                  $\sigma_{\overline{v_{\rm los}}}$ & 
                  $\langle \overline{\log{g}}-\log{g_{\odot}\rangle}$ & 
                  $\sigma_{\overline{\log{g}}}$ &
                  $\langle \overline{\rm [Fe/H]}-{\rm [Fe/H]}_{\odot}\rangle$ & 
                  $\sigma_{\overline{\rm [Fe/H]}}$ \\
      & (km s$^{-1}$) &(km s$^{-1}$) & (dex) & (dex) & (dex) &  (dex)\\
      \hline
      LMCC-b &  $-2.36$ & 0.18  & $0.01$ & 0.04 & $-0.18$ & 0.02 \\
      LMCC-r &  $-2.69$ & 0.17  & $0.02$ & 0.04 & $-0.18$ & 0.02 \\
      LMC1-b &  $-0.53$ & 0.22  & $-0.01$ & 0.04 & $-0.20$ & 0.02 \\
      LMC1-r &  $-0.35$ & 0.22 & $0.00$ & 0.04 & $-0.19$ & 0.02 \\
     \hline
    \end{tabular}
     \\
    \raggedright
    $^{a}$ library of \citet{Lee08a,Lee08b}\\
  \label{tab:twi_offsets}
\end{table*}
\begin{table*}
\begin{center}
\caption{Stellar parameters of the sample in the LMC bar}
\begin{tabular}{cccccc}
\hline\hline
ID & $S/N$ & $v_{\rm los}$ & $T_{\rm eff}$ & $\log{[g/(\rm cm\ s^{-2})]}$ & [Fe/H] \\
 &  & (km s$^{-1}$) & (K) & (dex) & (dex) \\
\hline
LMCC-b001 & 11.8 & $271.2\pm0.3$ & $4250\pm100$ & $1.20\pm0.09$ & $-0.63\pm0.05$ \\
LMCC-b002 & 13.3 & $259.3\pm0.3$ & $4085\pm75$ & $1.00\pm0.07$ & $-0.54\pm0.05$ \\
LMCC-b003 & 15.1 & $283.7\pm0.3$ & $4295\pm99$ & $1.15\pm0.07$ & $-0.73\pm0.04$ \\
LMCC-b004 & 13.3 & $223.3\pm0.3$ & $4341\pm106$ & $1.33\pm0.08$ & $-0.62\pm0.05$ \\
LMCC-b005 & 13.8 & $315.7\pm0.4$ & $4822\pm161$ & $1.11\pm0.18$ & $-1.50\pm0.05$ \\
$\cdots$ & & & & & \\
LMC1-b002 & 10.3 & $303.7\pm0.3$ & $4134\pm106$ & $0.84\pm0.08$ & $-0.65\pm0.05$ \\
LMC1-b003 & 9.6 & $278.4\pm0.4$ & $4250\pm112$ & $0.95\pm0.09$ & $-0.75\pm0.05$ \\
LMC1-b004 & 11.1 & $243.7\pm0.4$ & $4330\pm114$ & $1.26\pm0.09$ & $-0.65\pm0.05$ \\
LMC1-b005 & 9.9 & $268.2\pm0.4$ & $4758\pm199$ & $1.83\pm0.14$ & $-0.48\pm0.05$ \\
LMC1-b006 & 12.3 & $277.7\pm0.4$ & $4430\pm161$ & $1.47\pm0.09$ & $-0.72\pm0.05$ \\
$\cdots$ & & & & & \\
\hline
\end{tabular}
\end{center}
\label{tab:param}
\end{table*}

\begin{table*}
\begin{center}
\caption{Metallicity comparisons for 11 stars from the LMCC sample}
\begin{tabular}{cccccccc}
\hline\hline
ID & $S/N$ & $T_{\rm eff}$ & $\log{g}$ & [Fe/H]$_{\rm BA}$ & [Fe/H]$_{\rm EW}$ & [FeI/H]$_{\rm V13}$ & [Fe/H]$_{\rm C05}$ \\
   &  & (K) & (dex) & (dex) & (dex) & (dex) & (dex) \\
\hline
LMCC-b027 & 11.5 & $4886\pm174$ & $2.10\pm0.16$ & $-0.69\pm0.06$ & $-1.34\pm0.40$ &$-0.38\pm0.13$ & $-0.43\pm0.14$ \\
LMCC-b075 & 13.6 & $4278\pm140$ & $0.90\pm0.08$ & $-0.82\pm0.05$ & $-0.26\pm0.20$ &$-0.56\pm0.13$ & $-0.53\pm0.13$ \\
LMCC-b084 & 12.1 & $5107\pm201$ & $1.25\pm0.18$ & $-0.49\pm0.05$ & $-0.30\pm0.50$ &$-0.24\pm0.11$ & $-0.48\pm0.14$ \\
LMCC-b085 & 10.4 & $4922\pm177$ & $3.22\pm0.12$ & $-0.44\pm0.07$ & $-0.74\pm0.33$ &$-0.83\pm0.10$ & $-0.50\pm0.13$ \\
LMCC-b094 & 15.9 & $4951\pm168$ & $1.25\pm0.18$ & $-1.77\pm0.05$ & $-2.06\pm0.37$ &$-1.42\pm0.16$ & $-1.55\pm0.10$ \\
LMCC-b106 & 10.4 & $4237\pm123$ & $0.75\pm0.09$ & $-0.77\pm0.05$ & $-1.38\pm0.34$ &$-0.40\pm0.17$ & $-0.44\pm0.14$ \\
LMCC-b110 & 11.0 & $4154\pm 81$ & $1.07\pm0.08$ & $-0.59\pm0.05$ & $-0.94\pm0.24$ &$-0.23\pm0.22$ & $-0.16\pm0.14$ \\
LMCC-r043 & 17.6 & $4298\pm100$ & $1.46\pm0.08$ & $-0.47\pm0.05$ & $-0.62\pm0.21$ &$-0.69\pm0.12$ & $-0.20\pm0.14$ \\
LMCC-r064 & 19.9 & $4397\pm106$ & $1.01\pm0.08$ & $-0.93\pm0.04$ & $-0.88\pm0.29$ &$-0.58\pm0.11$ & $-1.42\pm0.13$ \\
LMCC-r068 & 18.3 & $4870\pm169$ & $0.84\pm0.13$ & $-0.50\pm0.04$ & $-0.46\pm0.41$ &$-0.25\pm0.12$ & $-0.33\pm0.14$ \\
LMCC-r078 & 17.2 & $4952\pm180$ & $0.73\pm0.12$ & $-0.45\pm0.04$ & $-0.36\pm0.45$ &$-0.12\pm0.11$ & $-0.30\pm0.14$ \\
\hline
\end{tabular}
\end{center}
\label{tab:comp}
\end{table*}

\begin{table*}
  \scriptsize
  \centering
    \caption{Kinematics versus metallicity for 239 stars}
    \begin{tabular}{cccc}
      \hline\hline
      [Fe/H] bin & $N_{\rm star}$ & $\overline{v_{\rm los}}$ & $\sigma_v$ \\ 
         & & (km s$^{-1}$) & (km s$^{-1}$)\\
      \hline
	$[-2.09, -1.10]$ & 23 & $261.8\pm6.7$ & $31.2\pm4.3$ \\
	$[-1.10, -0.80]$ & 58 & $251.1\pm4.3$ & $32.7\pm2.8$ \\
	$[-0.80, -0.60]$ & 99 & $262.3\pm2.8$ & $27.4\pm1.8$ \\
	$[-0.60, -0.38]$ & 59 & $258.2\pm2.5$ & $18.7\pm1.9$ \\
      \hline
    \end{tabular}
  \label{tab:kinematics}
\end{table*}

\newpage

\end{document}